\documentclass[
    aps,
    physrev,
    reprint,
    twocolumn,
    floatfix
]{revtex4-2}
\usepackage{amsmath}
\usepackage{amssymb}
\usepackage{bm}
\usepackage{graphicx}
\usepackage{mathtools}
\usepackage{braket}
\usepackage{tikz}
\usetikzlibrary{arrows.meta}
\usepackage{enumitem}
\usepackage{comment}
\usepackage[percent]{overpic}

\newcommand{\rL}{\mathrm{L}}
\newcommand{\rR}{\mathrm{R}}
\newcommand{\re}{\mathrm{e}}
\newcommand{\rh}{\mathrm{h}}

\newcommand{\rS}{\mathrm{S}}
\newcommand{\rF}{\mathrm{F}}
\newcommand{\rB}{\mathrm{B}}
\newcommand{\rd}{\mathrm{d}}
\newcommand{\cT}{\mathcal{T}}
\newcommand{\cR}{\mathcal{R}}
\newcommand{\bj}{\bar{j}}
\newcommand{\bp}{\bar{p}}

\newcommand{\bq}{\bar{q}}
\newcommand{\je}{J_\mathrm{ele}^{(\mathrm{DC})}}
\newcommand{\jh}{J_\mathrm{heat}^{(\mathrm{DC})}}

\begin{document}

\title{
    Electric and Heat Transport in One-Dimensional SNS Junctions with Energy-Dependent Transmission
}

\author{Kai Tsutsumi}
\affiliation{Department of Electronic and Physical Systems, Waseda University, Tokyo 169-8555, Japan}

\author{Shun Uchino}
\affiliation{Department of Electronic and Physical Systems, Waseda University, Tokyo 169-8555, Japan}
\affiliation{Department of Materials Science, Waseda University, Tokyo 169-8555, Japan}

\date{\today}

\begin{abstract}
   We investigate the phase-independent dc components of the electric and heat currents through a one-dimensional superconductor-normal-metal-superconductor (SNS) junction subject to chemical-potential and temperature differences between two superconducting reservoirs. Using a scattering approach based on the Bogoliubov--de Gennes equations, we incorporate multiple Andreev reflections while retaining the energy dependence of transmission through the normal region. We compare energy-independent, quantum-dot-type, and quantum-point-contact transmission profiles. At low bias, both currents exhibit nonlinear features associated with the superconducting gap and multiple Andreev reflections. Their responses to a temperature difference depend remarkably on the transmission profile. At a small finite bias, the electric current depends weakly on the temperature difference for the energy-independent and quantum-dot-type profiles, whereas the quantum-point-contact profile produces a pronounced nonlinear response. The heat current depends nonlinearly on the temperature difference for all three profiles over the parameter range investigated. These results highlight the role of the transmission profile in
   nonlinear electric and heat transport through SNS junctions
   and provide a microscopic reference for related transport
   studies in mesoscopic fermionic superfluid junctions.
\end{abstract}

\maketitle

\section{Introduction}
\label{sec:introduction}

Superconductor-normal-metal-superconductor (SNS) junctions
provide a basic playground for studying the interplay between
superconducting coherence and mesoscopic transport~\cite{Nazarov}.
They exhibit a variety of phenomena, including the superconducting proximity effect, Josephson currents, and nonequilibrium transport mediated
by Andreev reflection~\cite{Zagoskin}. 
Under a finite voltage bias, repeated electron-hole conversion
at the two normal-superconductor interfaces give rise to multiple
Andreev reflections (MARs), which produce characteristic subgap
structures in the current-voltage characteristics~\cite{Klapwijk,Gunsenheimer,AB}.
Their heat transport also reveals how superconducting coherence modifies
the transmission of energy-carrying quasiparticles~\cite{Maki,Zhao2003,Zhao2004}.
Beyond their fundamental interest, semiconductor-based weak links enable
electrostatic control of the Josephson coupling and have been used
to realize gate-tunable superconducting qubits~\cite{Larsen}.
Understanding electric and heat transport through such junctions
is therefore relevant to both nonequilibrium superconductivity~\cite{Nazarov2012} and 
superconducting device physics.

The role of the energy-dependent
 transmission is particularly clear
in normal-state thermoelectric transport. For a normal two-terminal junction close to equilibrium,
the electric and heat currents driven
by a chemical-potential difference $V=\mu_\rL-\mu_\rR$ and a temperature
difference $\Delta T=T_\rL-T_\rR$ can be written as~\cite{Houten}
\begin{equation}
    \begin{bmatrix}
        \je \\
        \jh
    \end{bmatrix}
    =
    \begin{bmatrix}
        G & L \\
        M & K
    \end{bmatrix}
    \begin{bmatrix}
        V / e \\
        \Delta T
    \end{bmatrix},
\label{eq:thermoelectric_transport}
\end{equation}
where $e$ is the elementary charge.
Here, $G$ and $K$ describe the direct electrical and
thermal responses, respectively, while $L$ and $M$
describe the thermoelectric cross response.
Within the Landauer-B\"{u}ttiker formalism~\cite{Datta},
these coefficients involve different energy moments of
the transmission probability ${\cal T}(E)$.
In particular, the cross coefficients depend on
the asymmetry of transmission above and below the equilibrium chemical potential. An energy-independent transmission gives $L=M=0$ in the usual wide-band limit,
whereas an asymmetric energy-dependent transmission can generate a finite thermoelectric response. Energy dependence alone is not sufficient: a symmetric transmission profile can still yield vanishing cross coefficients.
Thus, the value of the transmission at the Fermi level does not generally determine the thermoelectric properties of a junction.

In superconducting junctions, a widely used simplification is nevertheless to describe the normal region by energy-independent scattering parameters. Examples include tunnel-Hamiltonian models with energy-independent coupling and a wide-band normal-state density of states~\cite{Cuevas}, as well as short-contact implementations of quasiclassical Green-function and circuit theories in which normal-state transmission probabilities are evaluated at the Fermi level~\cite{Nazarov1999,Gunsenheimer,Bratus,Virtanen}. These treatments retain the energy dependence arising from superconductivity itself, yet neglect the additional energy dependence associated with scattering in the normal region.
This approximation is appropriate when the normal-state scattering properties vary negligibly over the relevant transport energy window. It can, however, miss important effects when that window contains a narrow transmission resonance or a pronounced transmission threshold.

Energy-dependent superconducting transport has been investigated in resonant-level models and in theories of 
superconducting thermoelectricity~\cite{Levy,Johansson,Martin,Pershoguba}.
These studies motivate examining how different transmission profiles affect electric and heat currents in the presence of both chemical-potential and temperature differences.
At finite bias, MAR couples scattering processes at different energies, so that a quasiparticle repeatedly traversing the normal region samples its transmission over a sequence of bias-shifted energies~\cite{AB,Hofmann}. The influence of an energy-dependent transmission therefore involves its interplay with superconducting conversion processes, rather than simply a change in the normal-state conductance. A joint analysis of electric and heat transport is needed to determine how this interplay affects their respective bias and temperature responses beyond linear response.

Experiments with ultracold Fermi gases provide a complementary setting for
exploring these questions through measurements of particle and entropy transport with
tunable interactions and chemical potentials~\cite{Brantut,Krinner2015,Krinner,Amico}.
Gate-controlled measurements have demonstrated an interaction-assisted reversal of
the thermoelectric particle current, highlighting the competition between energy-selective
transport through the channel and the thermodynamic response of the reservoirs~\cite{Samuel}.
Experiments with superfluid reservoirs have also revealed strongly nonlinear particle and 
entropy currents driven by chemical-potential and temperature differences~\cite{Fabritius,Mohan}.
Near unitarity~\cite{Zwerger},
the magnitudes of the advective and diffusive currents were found to depend strongly on
the channel potential and confinement, whereas the entropy advectively transported per particle
was largely insensitive to these channel details over the parameter range explored.
These contrasting observations motivate identifying which aspects of nonlinear transport are sensitive to the transmission properties of the junction.

In this work, we calculate the phase-independent dc components of the electric and heat currents through a
one-dimensional SNS junction subject to chemical-potential
and temperature differences. Using a scattering formulation
based on the Bogoliubov--de Gennes equations, we incorporate
MAR while retaining the energy dependence of transmission through the normal region. 
We compare three representative transmission profiles:
an energy-independent profile, a Lorentzian quantum-dot-type profile, and a smooth step-like quantum-point-contact (QPC) profile. 
By calculating both currents as functions of the chemical-potential bias and temperature
difference, we examine how flat, resonant, and threshold-like transmission characteristics
affect nonlinear transport within a common framework. Our focus is on the phase-independent dc response, rather than on current-phase relations or ac Josephson harmonics. This comparison allows us to examine 
the influence of the transmission profile within
a mean-field scattering description and provides a basis for understanding
nonlinear particle and thermal transport in fermionic superfluid junctions.

For the energy-independent transmission, the calculated electric current reproduces the established MAR result~\cite{AB}. At a small finite bias, we find that the electric current depends only weakly on the temperature difference for the energy-independent and centered quantum-dot-like profiles, whereas the quantum-point-contact profile produces a pronounced nonlinear temperature response.
The heat current exhibits nonlinear temperature-difference dependence for all three profiles over the parameter range investigated. These contrasting responses highlight the role of the transmission profile in determining both electric and heat transport and show why the two currents should be considered together when characterizing nonlinear transport through SNS junctions.

This paper is organized as follows. In Sec.~\ref{sec:model_and_scattering_formalism}, we introduce
the model and the scattering formulation, including the recurrence relations for the scattering amplitudes.
 Section~\ref{sec:currents} derives the expressions for the dc electric and heat currents. 
 In Sec.~\ref{sec:results}, we present numerical results for the three transmission profiles and discuss their bias and temperature-difference dependence. 
  Section~\ref{sec:conclusion} summarizes our conclusions.
  Appendix~\ref{app:derivation_recurrence} provides a detailed derivation of the recurrence relation. Appendix~\ref{app:numerical_solution_of_the_recurrence_relation} describes the numerical procedure used to solve it, and Appendix~\ref{app:derivation_currents}
  presents the derivations of the electric- and heat-current expressions.

\section{Model and Scattering Formalism}
\label{sec:model_and_scattering_formalism}

\subsection{Model}

\begin{figure}[htb]
    \centering
    \begin{tikzpicture}
    \draw (0, 2) -- (7.5, 2);
    \draw (2.5, 2) -- (2.5, 0);
    \draw (5, 2) -- (5, 0);
    \draw (0, 0) -- (7.5, 0);
    
    \node at (1.25, 1.2){\Large $\text{S}_\text{L}$};
    \node at (1.25, 0.5) {$T_\text{L}, \theta_\text{L}, \mu_\text{L}$};
    \node at (3.75, 1.2){\Large N};
    \node at (3.75, 0.5) {$\mathcal{T}(E)$};
    \node at (6.25, 1.2){\Large $\text{S}_\text{R}$};
    \node at (6.25, 0.5) {$T_\text{R}, \theta_\text{R}, \mu_\text{R}$};
\end{tikzpicture}
    \caption{Schematic of the one-dimensional SNS junction 
    considered in this work. Two superconducting reservoirs, $\rS_\rL$ and
    $\rS_\rR$, are connected by a normal region with an energy-dependent transmission probability ${\cal T}(E)$.}
    \label{fig:sns_setup}
\end{figure}
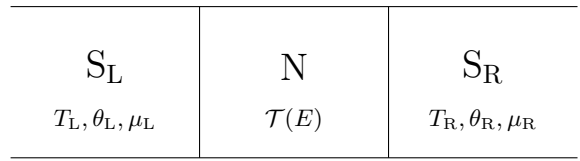

We consider a one-dimensional SNS junction
consisting of two superconducting reservoirs, $\rS_\rL$ and $\rS_\rR$,
connected by a normal region, as illustrated in Fig.~\ref{fig:sns_setup}.
The reservoirs are characterized by their temperatures $T_j$,
chemical potentials $\mu_j$, and constant superconducting phases
$\theta_j$, where $j = \rL,\rR$. We define the temperature difference
$\Delta T$, relative phase $\varphi$, and chemical-potential bias $V$ by
\begin{equation}
    T_\rL = T_\rR + \Delta T
    , \quad
    \theta_\rL = \theta_\rR + \varphi
    , \quad
    \mu_\rL = \mu_\rR + V
    .
\end{equation}
Here, $V$ has units of energy, and therefore the corresponding voltage difference is $V/e$.
Since we focus on the phase-independent dc components of the currents, we set  $\varphi = 0$.
This choice fixes only the constant phase offset;
the bias-induced time dependence is retained through the sideband
structure introduced below.

For notational convenience, we use $j = \rL,\rR$ to label the two sides of
the junction and $p = \re, \rh$ to label electrons and holes.
A bar denotes the opposite side or particle type.
We also introduce the sign factor $\zeta_{ab}$:
\begin{equation}
\begin{gathered}
        \bj
        =
        \begin{cases}
                \rR     & : j = \rL
            ,\\
                \rL     & : j = \rR
        \end{cases}
        ;
    \\
        \bp
        =
        \begin{cases}
                \rh     & : p = \re
            ,\\
                \re     & : p = \rh
        \end{cases}
        ;
    \\
        \zeta_{ab}
        =
        \begin{cases}
                +1  & : a = b 
            ,\\
                -1  & : a \neq b
        \end{cases}
        .
\end{gathered}
\end{equation}

Electron scattering in the normal region is described
by the unitary scattering matrix
\begin{equation}
    S
    =
    \begin{bmatrix}
            r_\rL^{(\re)}(E)
        &
            t_\rR^{(\re)}(E)
        \\
            t_\rL^{(\re)}(E)
        &
            r_\rR^{(\re)}(E)
    \end{bmatrix},
\end{equation}
where $r_j^{(\re)}(E)$ and $t_j^{(\re)}(E)$ are the reflection
and transmission amplitudes for an electron incident from side $j$.
The corresponding reflection and transmission probabilities are
denoted by $\cR(E)$ and  $\cT(E)$, respectively. 
Unitarity imposes the following conditions:
\begin{equation}
\begin{gathered}
        r_j^{(\rh)}(-E) \cdot r_j^{(\re)}(E) 
        =
        \cR(E)
    , \\
        t_j^{(\rh)}(-E) \cdot t_j^{(\re)}(E)
        =
        \cT(E)
    , \\
        \cR(E) + \cT(E) = 1
    , \\
        r_\rL^{(\rh)}(-E) \cdot t_\rR^{(\re)}(E)
        +
        t_\rL^{(\rh)}(-E) \cdot r_\rR^{(\re)}(E)
        =
        0
    .
\end{gathered}
\end{equation}
The normal-state scattering amplitudes for holes are 
related to those for electrons by
\begin{equation}
    r_j^{(\rh)}(-E) = {r_j^{(\re)}(E)}^*
    , \quad
    t_j^{(\rh)}(-E) = {t_j^{(\re)}(E)}^*
    .
    \label{eq:particle-hole}
\end{equation}
Equation~\eqref{eq:particle-hole} relates the electron and hole
scattering amplitudes; it does not require the transmission
probability $\cT(E)$ to be an even function of energy.

For simplicity, we restrict the normal-state scattering amplitudes
to be real and impose
\begin{align}
\begin{gathered}
        r_j^{(p)}(\zeta_{p \re} E), t_j^{(p)}(\zeta_{p \re} E)
        \in
        \mathbb{R};
    \\
        \begin{aligned}
                r_{\rR}^{(p)}(\zeta_{p \re} E)
            &
                =
                -r_{\rL}^{(p)}(\zeta_{p \re} E);
            \\
                t_{\rR}^{(p)}(\zeta_{p \re} E)
            &
                =
                t_{\rL}^{(p)}(\zeta_{p \re} E).
        \end{aligned}        
\end{gathered}
\end{align}

At the NS interfaces, we define the Andreev-reflection amplitudes $r_j^{(p \bp)}(\zeta_{p \re} E)$ and the injection amplitudes from the superconducting region to the normal region $t_j^{(p \bp)}(\zeta_{p \re} E)$ as
\begin{align}
\begin{aligned}
        r_j^{(p \bp)}(\zeta _{p \re} E)
    &
        =
        \frac{v_0(E)}{u_0(E)}
        e^{-\zeta_{p \re} i \theta_j}
        , \quad
    \\
        t_j^{(p \bp)}(\zeta _{p \re} E)
    &
        =
        \left( u_0(E) - \frac{v_0(E)^2}{u_0(E)} \right)    
        e^{\zeta_{p \re} i \theta_j / 2}.
\end{aligned}
\end{align}
For example, $r_\rL^{(\re \rh)}(E)$ denotes the amplitude for an electron incident on the $\mathrm{NS_\rL}$ interface to undergo Andreev reflection as a hole. The quantities appearing in these amplitudes are defined by
\begin{align}
\begin{gathered}
        \xi(E)
        =
        \begin{cases}
                \sqrt{E^2 - \Delta^2}     & : \Delta < E,
            \\
                i \sqrt{\Delta^2 - E^2}   & : |E| < \Delta,
            \\
                -\sqrt{E^2 - \Delta^2}    & : E < -\Delta
        \end{cases}
        ,
    \\
        u_0(E)
        =
        \sqrt{\frac{1}{2} \left( 1 + \frac{\xi(E)}{E} \right)}
        ,
    \\
        v_0(E)
        =
        \sqrt{\frac{1}{2} \left( 1 - \frac{\xi(E)}{E} \right)}
        .
\end{gathered}
\end{align}
Here, $\Delta$ denotes the common superconducting gap
magnitude entering the interface scattering amplitudes.

\subsection{Scattering Relations}
\label{subsec:scattering_relations}

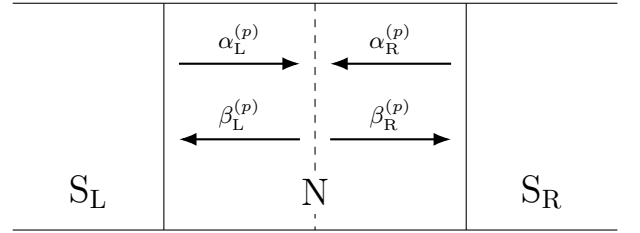
\begin{figure}[htb]
    \centering
    \begin{tikzpicture}
    \draw (0, 3) -- (8, 3);
    \draw (2, 3) -- (2, 0);
    \draw (6, 3) -- (6, 0);
    \draw (0, 0) -- (8, 0);
    \draw[dashed] (4, 3) -- (4, 0);
    
    \draw[-Latex, thick] (2.2, 2.2) -- node[above]{$\alpha_\mathrm{L}^{(p)}$} (3.8, 2.2);
    \draw[Latex-, thick] (4.2, 2.2) -- node[above]{$\alpha_\mathrm{R}^{(p)}$} (5.8, 2.2);
    \draw[Latex-, thick] (2.2, 1.2) -- node[above]{$\beta_\mathrm{L}^{(p)}$} (3.8, 1.2);
    \draw[-Latex, thick] (4.2, 1.2) -- node[above]{$\beta_\mathrm{R}^{(p)}$} (5.8, 1.2);

    \node at (1, 0.5){\Large $\text{S}_\text{L}$};
    \node at (7, 0.5){\Large $\text{S}_\text{R}$};
    \node[fill = white, inner sep = 3pt] at (4, 0.5){\Large $\text{N}$};
\end{tikzpicture}
    \caption{The scattering amplitudes in the normal region. The amplitudes $\alpha^{(p)}_j$ and $\beta^{(p)}_j$ describe modes propagating toward and away from the normal-region scatterer, respectively. The arrows indicate the propagation directions.}
    \label{fig:def_coefficients}
\end{figure}
The wave-function amplitudes in the normal region are
defined as illustrated in Fig.~\ref{fig:def_coefficients}.
The amplitudes $\alpha_j^{(p)}$ describe modes
propagating toward the normal-region scatterer, whereas
$\beta_j^{(p)}$ describe modes propagating away from
it and toward the superconducting reservoirs.

Consider a scattering state generated by a particle of
type $q\,(=\re,\rh)$ with energy
$\zeta_{q\re}E$, injected from the superconducting reservoir
$\rS_k$, where $k=\rL,\rR$. We keep the injection channel $(k,q)$
and the injection-energy parameter $E$ fixed in the following
derivation and suppress their labels in the scattering
amplitudes.

At finite bias, the scattering relations couple amplitudes
at energies shifted by integer multiples of $V$.
We label these sidebands by $n\in \mathbb{Z}$.
Combining scattering in the normal region with Andreev reflection
and injection at the NS interfaces, we obtain
\begin{widetext}
    \begin{align}
    \begin{aligned}
        &
            \beta_j^{(p)}\big( \zeta_{p \re}(E + nV) \big)
        \\ &
            \begin{aligned}
                    ={}
                &
                    r_j^{(p)}\Big(
                        \zeta_{p \re}
                        \big( E + (n + \zeta_{p \re}\delta_{j \rL}) V \big)
                    \Big)
                    \bigg(
                        \delta_{j, k} \delta_{p, q} \delta_{n, 0}
                        \cdot
                        \iota(E + nV)
                        +
                        r_j^{(\bp p)}\big( \zeta_{p \re}(E + nV) \big)
                        \beta_j^{(\bp)}\big( \zeta_{\bp \re}(E + nV) \big)
                    \bigg)
                \\ &
                    +
                    t_{\bj}^{(p)}\Big(
                        \zeta_{p \re}\big( E + (n + \zeta_{p \re}\delta_{j \rL}) V \big)
                    \Big)
                    \bigg(
                        \delta_{\bj, k} \delta_{p, q}
                        \delta_{n + \zeta_{p \re} \zeta_{j \rL}, 0}
                        \cdot
                        \iota \big( E + (n + \zeta_{j \rL} \zeta_{p \re}) V \big)
                \\ &
                        +
                        r_{\bar{j}}^{(\bp p)}\Big(
                            \zeta_{p \re}
                            \big( E + (n + \zeta_{j \rL} \zeta_{p \re}) V \big)
                        \Big)
                        \beta_{\bj}^{(\bp)}\Big(
                            \zeta_{\bp \re}\big( E + (n + \zeta_{j \rL} \zeta_{p \re}) V \big)
                        \Big)
                    \bigg)
            \end{aligned}
    \label{eq:scattering-relation_full}        
    \end{aligned}
    \end{align}
    For compactness, we rewrite this relation as
    \begin{align}
        \beta_j^{(p)}(n)
        ={}
        \delta_{j, k} \delta_{p, q} \delta_{n, 0}
        I_j^{(p)}(n)
        +
        A_j^{(\bp)}(n)
        \beta_j^{(\bp)}(n)
        +
        \delta_{\bj, k} \delta_{p, q} \delta_{n \pm 1, 0}
        J_{\bj}^{(p)}(n \pm 1)
        +
        B_{\bj}^{(\bp)}(n \pm 1)
        \beta_{\bj}^{(\bp)}(n \pm 1).
    \label{eq:scattering-relation_compact}
    \end{align}
\end{widetext}
Here, $I_j^{(p)}$ and $J_j^{(p)}$ are source coefficients
associated with injection from the reservoirs, while $A_j^{(p)}$
and $B_j^{(p)}$ combine normal-region scattering and Andreev-reflection amplitudes. These coefficients are
defined by comparison with 
Eq.~\eqref{eq:scattering-relation_full}.
The injection factor $\iota(E)$ appearing in Eq.~\eqref{eq:scattering-relation_full} is defined, following Refs.~\cite{AB,Chauvin}, as
\begin{equation}
    \iota(E)
    =
    \frac{
        u_0(E)^2 - v_0(E)^2
    }{
        u_0(E)
    }.
\end{equation}

For each fixed injection channel $(k,q)$,
Eq.~\eqref{eq:scattering-relation_compact} represents four 
coupled equations corresponding to the four choices of $(j,p)$.
Considering all four injection channels therefore gives 16
scalar equations.

The abbreviated energy arguments in Eq.~\eqref{eq:scattering-relation_compact} are defined by
\begin{equation}
\begin{aligned}
        \zeta_{p \re}(E + nV)
    &
        \to n
    , \\
        \zeta_{p \re}\big( E + (n + m \zeta_{p \re} \zeta_{j \rL}) V \big)
    &
        \to n \pm m
    , \\
        \zeta_{p \re}\big( E + (n - m \zeta_{p \re} \zeta_{j \rL}) V \big)
    &
        \to n \mp m
\end{aligned}
\label{eq:sideband_index_notation}
\end{equation}
where $m=1,2,3,\cdots$.
Throughout the following derivation, the upper signs  are used for $(j,p)=(\rL,\re)$ and $(\rR,\rh)$, and the lower signs for $(j,p)=(\rL,\rh)$ and $(\rR,\re)$.

For later use, we note that the notation introduced in
Eq.~\eqref{eq:sideband_index_notation} obeys
\begin{equation}
\begin{aligned}
        \begin{rcases}
                p \to \bp
            , \\
                j \to \bj
        \end{rcases}
    &
        \Rightarrow
        \begin{cases}
                n \pm m \to n \mp m
            , \\
                n \mp m \to n \pm m
        \end{cases}
    ; \\
        n \to n \pm m'
    &
        \Rightarrow
        \begin{cases}
                n \pm m \to n \pm (m + m')
            , \\
                n \mp m \to n \mp (m - m')
        \end{cases}
    ; \\
        n \to n \mp m'
    &
        \Rightarrow
        \begin{cases}
                n \pm m \to n \pm (m - m')
            , \\
                n \mp m \to n \mp (m + m')
        \end{cases}
\end{aligned}
\label{eq:sideband_index_rules}
\end{equation}
where $m'=1,2,3,\cdots$. In particular, for $m'=\delta_{j,\rL}$, we have
\begin{equation}
    j \to \bj
    \Rightarrow
    \begin{cases}
            n \pm \delta_{j, \rL}
            \to
            n \mp (1 - \delta_{j, \rL})
        , \\
            n \mp \delta_{j, \rL}
            \to
            n \pm (1 - \delta_{j, \rL})
    \end{cases}.
\end{equation}

\subsection{Recurrence relation for a specified injection channel}
\label{subsec:recurrence-relation}

We now specialize to the injection channel specified by 
$j=k$ and $p=q$. Combining 
Eq.~\eqref{eq:scattering-relation_compact} with the equations obtained by the appropriate replacements $j\to\bj$ and $p\to\bp$, 
we eliminate all scattering amplitudes other than $\beta_j^{(p)}$.
This yields the three-term recurrence relation
\begin{equation}
\begin{aligned}
        \delta_{n,0} I_j^{(p)}(n)
        ={}
    &
        R_j^{(p)}(n \pm 2) \beta_j^{(p)}(n \pm 2)
    \\ &
        +
        S_j^{(p)}(n) \beta_j^{(p)}(n)
    \\ &
        +
        T_j^{(p)}(n \mp 2) \beta_j^{(p)}(n \mp 2),
\end{aligned}
\label{eq:recurrence-formula}
\end{equation}
The auxiliary coefficients $R_j^{(p)}$, $S_j^{(p)}$, $T_j^{(p)}$
are given explicitly in Appendix~\ref{app:derivation_recurrence}.
Equation~\eqref{eq:recurrence-formula} is equivalent to the recurrence relation in Ref.~\cite{Hofmann}.

We solve Eq.~\eqref{eq:recurrence-formula} numerically 
using the procedure described in Appendix~\ref{app:numerical_solution_of_the_recurrence_relation}.
Once $\beta_j^{(p)}$ is known, the remaining amplitudes
on the same side of the junction follow from
\begin{align}
    &
        \begin{aligned}
                \beta_j^{(\bp)}(n)
                ={}
            &
                G_j^{(p)}(n \mp 2)
                \beta_j^{(p)}(n \mp 2)
            \\ &
                +
                H_j^{(p)}(n)
                \beta_j^{(p)}(n),
        \end{aligned}
\label{eq:beta-bp}
    \\ &
        \alpha_j^{(p)}(n)
        =
        r_j^{(\bp p)}(n)
        \beta_j^{(\bp)}(n)
        +
        \delta_{n,0}
        \frac{u_0(n)^2 - v_0(n)^2}{u_0(n)},
    \\ &
        \alpha_j^{(\bp)}(n)
        =
        r_j^{(p\bp)}(n)
        \beta_j^{(p)}(n).
\end{align}
The derivations 
 of Eqs.~\eqref{eq:recurrence-formula} and~\eqref{eq:beta-bp},
 together with the definitions of $G_j^{(p)}$ and $H_j^{(p)}$,
 are provided in Appendix~\ref{app:derivation_recurrence}.
Repeating this procedure for the four injection channels gives
the scattering amplitudes required to evaluate the currents.

\section{Currents}
\label{sec:currents}

In this section, we express the dc electric and heat currents
in terms of the scattering amplitudes obtained in Sec.~\ref{sec:model_and_scattering_formalism}.
Details of the derivations are provided in Appendix~\ref{app:derivation_currents}.

\subsection{Field-operator expansion}
For a fixed incoming channel $(k,q)$, the corresponding contribution
 to the Nambu field 
 can be expanded in scattering states as
\begin{align}
\begin{aligned}
        \begin{bmatrix}
                \hat{\psi}_{\uparrow k}^{(q)}(x, t)
            \\
                {\hat{\psi}_{\downarrow k}^{(q)}(x, t)}^\dagger
        \end{bmatrix}
        ={}
    &
        \frac{1}{\sqrt{h v_\rF}}
        \int_{-\infty}^\infty \rd E
            \sqrt{D(E - \delta_{k \rL} \zeta_{q \re}V)}
    \\ &
            \cdot
            \Psi_k^{(q)}(E, x, t)
            \hat{a}_{k \sigma_q}^{(q)}(\zeta_{q \re} E - \delta_{k \rL} V)
\end{aligned}
\label{eq:field-operator_expansion}
\end{align}
Here, the operator $\hat{a}_{k \sigma_q}^{(q)}(\zeta_{q \re} E)$ 
is the annihilation operator associated with the incoming mode,
with $\sigma_\re ={} \uparrow$ and $\sigma_\rh ={} \downarrow$.
The function $\Psi_k^{(q)}(E, x, t)$ is the corresponding
Nambu scattering wave function in the normal region;
its explicit form is given in Appendix~\ref{app:derivation_currents}.
We use $h=2\pi\hbar$, and $v_\rF$ denotes the normal-state Fermi velocity.

The dimensionless function $D(E)$ is the superconducting density of states normalized to its normal-state value:
\begin{align}
    D(E)
    =
    \begin{cases}
            \frac{|E|}{\sqrt{E^2 - \Delta^2}}
        &
            : |E|>\Delta,
        \\
            0
        &
            : |E|<\Delta
    \end{cases},
\end{align}
Equation~\eqref{eq:field-operator_expansion} represents
the contribution from the incoming channel $(k,q)$.
Contributions from all incoming channels are included when
evaluating the currents below.

\subsection{Electric current}
We define the electric current $J_\mathrm{ele}(x,t)$ as
\begin{align}
\begin{aligned}
    &
        J_\mathrm{ele}(x, t)
    \\ &
        =
        \frac{e \hbar}{2 i m}
        \sum_{k, q}
        \sum_{\sigma = \uparrow, \downarrow}
        \left\langle
            {\hat{\psi}_{\sigma k}^{(q)}{}}^\dagger
            \frac{\partial
                \hat{\psi}_{\sigma k}^{(q)}
            }{\partial x}
            -
            \frac{\partial
                {\hat{\psi}_{\sigma k}^{(q)}{}}^\dagger
            }{\partial x}
            \hat{\psi}_{\sigma k}^{(q)}
        \right\rangle.
\end{aligned}
\label{eq:def_electric-current}
\end{align}
Notice that we use the sign convention in
Eq.~\eqref{eq:def_electric-current},
in which the electric current is $e$ times
the particle current, with $e>0$.
Substituting the field-operator expansion \eqref{eq:field-operator_expansion} into Eq.~\eqref{eq:def_electric-current} and retaining only the time-independent dc component, we obtain
\begin{widetext}
\begin{align}
\begin{aligned}
        \je
        ={}
    &
        \frac{e}{h v_\rF}
        \sum_{p, q} \sum_n \sum_\gamma
        \zeta_{\gamma \alpha}
        \int_{-\infty}^\infty \rd E \
        \left[
            D(E - \zeta_{q \re}V)
            \left|
                \gamma_\rL^{(pq)}(E, n - \zeta_{p \re})
            \right|^2
            f_\rL\big( \zeta_{p \re} (E - \zeta_{q \re} V) \big)
        \right.
    \\ &
        \left.
            -
            D(E)
            \left|
                \gamma_\rR^{(pq)}(E, n)
            \right|^2
            f_\rR(\zeta_{p \re} E)
        \right].
\end{aligned}
\label{eq:electric-current}
\end{align}
\end{widetext}

The occupation factors are expressed 
in terms of the Fermi distribution of
reservoir $S_k$,
\begin{align}
    f_k(E)
    =
    \frac{1}{
        1
        +
        \exp{
        \left(
            \frac{E}{k_\rB T_k}
        \right)
        }
    },
\label{eq:fermi-distribution}
\end{align}
where $E$ is measured relative to the 
chemical potential of the right superconducting
reservoir $\mu_\rR$.

Note that the notation is generalized, for example, as $\gamma_j^{(p)}(n) \to \gamma_j^{(pq)}(E,n)$. In Sec.~\ref{subsec:recurrence-relation}, the injection energy $E$ and the injected particle type $q$ were fixed throughout and were therefore suppressed in the notation. Since they are not fixed in the present subsection, we make these indices and arguments explicit.

\subsection{Heat current}
The heat current $J_\mathrm{heat}(x,t)$ is given by
\begin{align}
\begin{aligned}
    &
        J_\mathrm{heat}(x, t)
    \\ &
        \begin{aligned}
                ={}
            &
                \frac{-\hbar^3}{4 i m^2} \sum_{k, q} \sum_\sigma
                \left\langle
                    \frac{\partial^2
                        {\hat{\psi}_{\sigma k}^{(q)}{}}^\dagger
                    }{\partial x^2}
                    \frac{\partial
                        \hat{\psi}_{\sigma k}^{(q)}
                    }{\partial x}
                    -
                    \frac{\partial
                        {\hat{\psi}_{\sigma k}^{(q)}{}}^\dagger
                    }{\partial x}
                    \frac{\partial^2
                        \hat{\psi}_{\sigma k}^{(q)}
                    }{\partial x^2}
                \right\rangle
            \\ &
                -
                \frac{\hbar}{2 i m}
                \sum_{k, q}
                \sum_\sigma
                \mu_k
                \left\langle
                    {\hat{\psi}_{\sigma k}^{(q)}{}}^\dagger
                    \frac{\partial
                        \hat{\psi}_{\sigma k}^{(q)}
                    }{\partial x}
                    -
                    \frac{\partial
                        {\hat{\psi}_{\sigma k}^{(q)}{}}^\dagger
                    }{\partial x}
                    \hat{\psi}_{\sigma k}^{(q)}
                \right\rangle.
        \end{aligned}
\end{aligned}
\label{eq:formalism_heat-current}
\end{align}
The first term is the kinetic-energy-current
contribution, while the second subtracts
the chemical-potential-weighted particle-current
contribution.

Substituting the field-operator expansion \eqref{eq:field-operator_expansion} into Eq.~\eqref{eq:formalism_heat-current} and retaining only the time-independent dc component, we obtain (see also Appendix~\ref{app:derivation_currents})
\begin{widetext}
\begin{align}
\begin{aligned}
        \jh
        ={}
    &
        \frac{1}{h v_\rF}
        \sum_{p, q} \sum_n \sum_\gamma
        \zeta_{\gamma \alpha}
        \zeta_{p \re}
        \int_{-\infty}^\infty \rd E \
        (E + nV)
        \cdot
        \left[
            D(E - \zeta_{q \re}V)
            \left|
                \gamma_\rL^{(pq)}(E, n - \zeta_{p \re})
            \right|^2
            f_\rL\big( \zeta_{p \re} (E - \zeta_{q \re} V) \big)
        \right.
    \\ &
        \left.
            -
            D(E)
            \left|
                \gamma_\rR^{(pq)}(E, n)
            \right|^2
            f_\rR(\zeta_{p \re} E)
        \right].
\end{aligned}
\label{eq:heat-current}
\end{align}
\end{widetext}

Note that here again we use the extended notation $\gamma_j^{(p)}(n) \to \gamma_j^{(pq)}(E,n)$.
We point out that the heat current obtained above
is consistent with one discussed in Ref.~\cite{Yamamoto}.

\section{Results}
\label{sec:results}

\subsection{Transmission Probabilities}
The formulation developed above retains the energy
dependence of the transmission probability in the 
normal region. We now use this formulation to compare
three representative transmission profiles:
an energy-independent profile, a quantum-dot-type Lorentzian profile, and a QPC
profile. The corresponding transmission probabilities
are
\begin{align}
        \cT_\mathrm{const}
    &
        =
        \cT_0
    ,\\
        \cT_\mathrm{Qdot}
    &
        =
        \frac{1}{
            1
            +
            \left(
                \frac{E - E_0}{\Gamma}
            \right)^2
        }
    ,\\
        \cT_\mathrm{QPC}
    &
        =
        \frac{1}{
            1
            +
            \exp{\left(
                -\frac{2\pi(E - E_*)}{\hbar \Omega}
            \right)}
        }.
\end{align}
Here, ${\cal T}_0$ is the constant transmission 
probability, $\Gamma$ is the half-width at half-maximum of the Lorentzian resonance, and $\hbar\Omega$
controls the width of the QPC transmission step.
The forms of $\cT_\mathrm{Qdot}$ and $\cT_\mathrm{QPC}$ are chosen with reference to Refs.~\cite{Nazarov} and \cite{Buttiker,Lebrat}, respectively.

We place both the resonance energy and the transmission
threshold at the mean chemical potential of the reservoirs,
\begin{equation}
    E_0 = E_* = \frac{\mu_\rL + \mu_\rR}{2},
    \label{eq:E0_and_E*}
\end{equation}
and vary $\cT_0$, $\Gamma$, and $\hbar \Omega$.  Increasing $\Gamma$ broadens the resonance 
without changing its peak transmission,
whereas increasing $\hbar\Omega$ makes the QPC
transmission step more gradual. The Lorentzian profile
is symmetric about $E_0$, while the QPC profile increases
monotonically through its threshold at $E_*$.

\begin{figure}[t]
    \centering

    \begin{overpic}[width = \columnwidth]{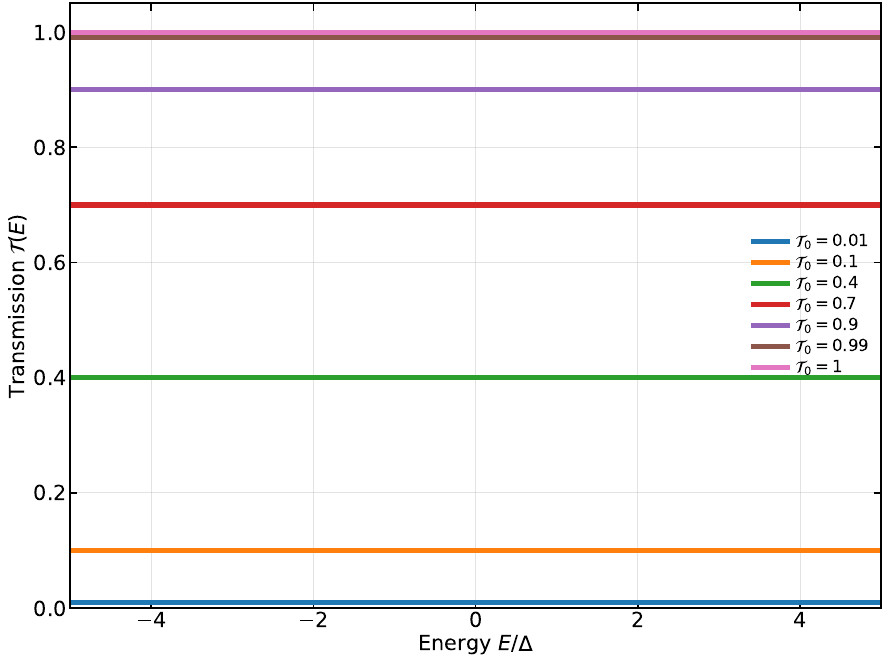}
        \put(5,80){\bfseries (a)}
    \end{overpic}
    
    \vspace{30pt}
    
    \begin{overpic}[width = \columnwidth]{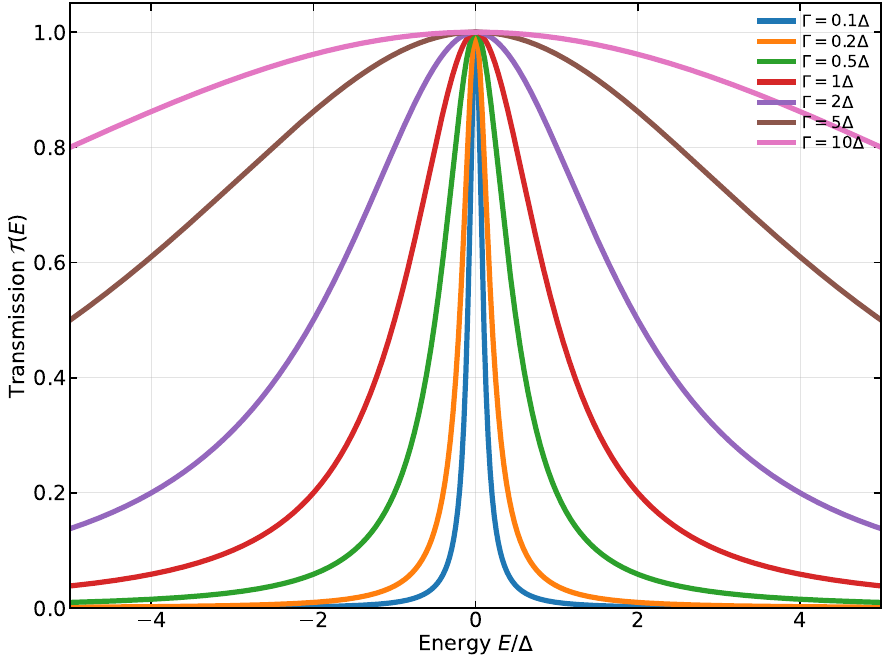}
        \put(5,80){\bfseries (b)}
    \end{overpic}
    
    \vspace{30pt}
    
    \begin{overpic}[width = \columnwidth]{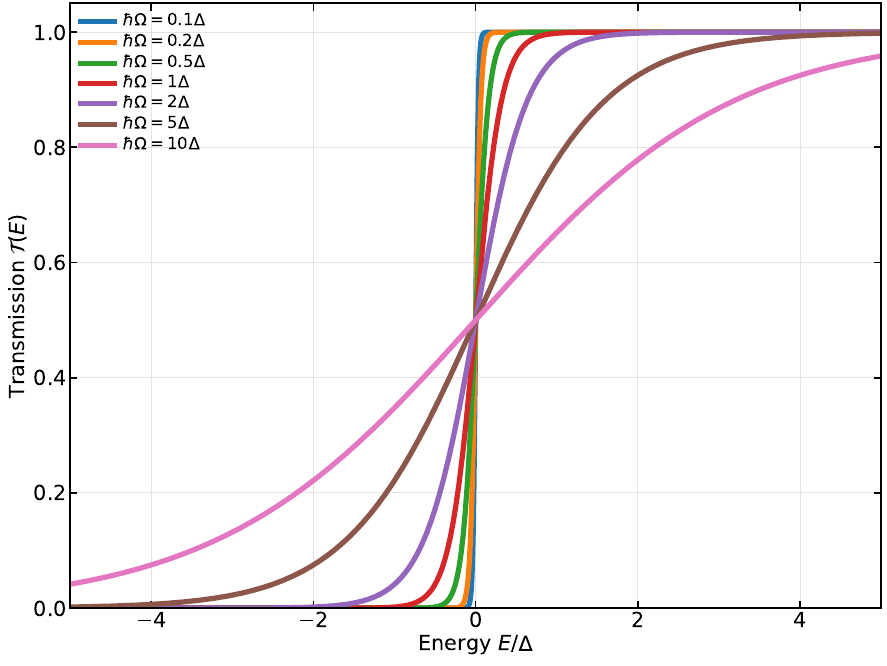}
        \put(5,80){\bfseries (c)}
    \end{overpic}

    \caption{
        The normal-region transmission probability as a function
        of energy:
        (a) energy-independent,
        (b) quantum-dot-type,
        and (c) QPC profiles.
        The legends give the  
        values of ${\cal T}_0$, $\Gamma$,
        and $\hbar\Omega$, respectively.
    }
    \label{fig:transmission}
\end{figure}

At fixed energy, the QPC transmission tends to zero
or unity in the limits $E_* \to \infty$ or $E_* \to -\infty$, respectively. Here, however, we use the choice
in Eq.~\eqref{eq:E0_and_E*} to examine the effect of a transmission threshold near the reservoir chemical potentials.
The three  transmission profiles are illustrated in Fig.~\ref{fig:transmission}.

\subsection{Bias dependence}

\begin{figure*}[t]
    \centering

    \begin{overpic}[width=0.32\textwidth]{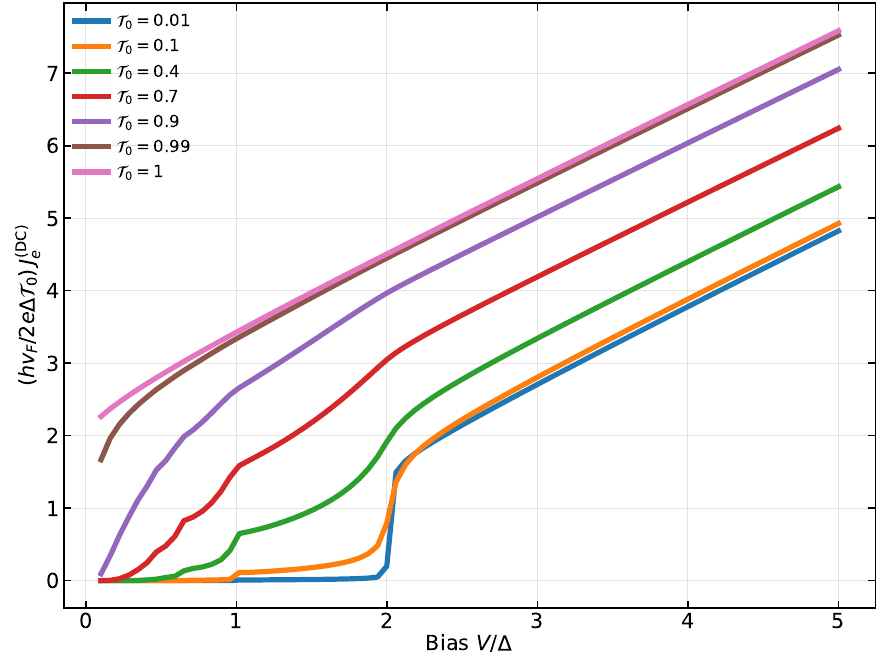}
        \put(3,80){\bfseries (a)}
    \end{overpic}    
    \hfill    
    \begin{overpic}[width=0.32\textwidth]{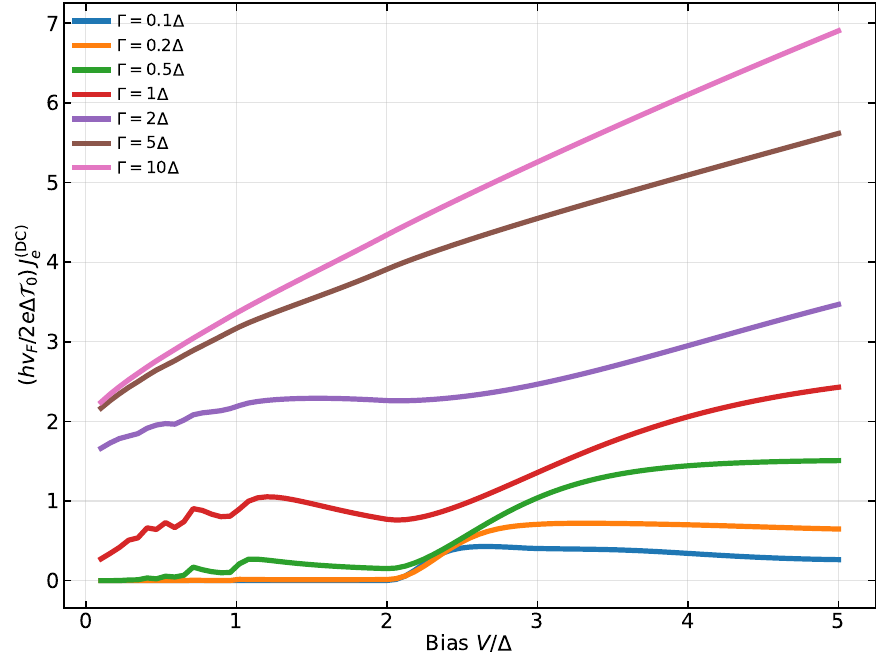}
        \put(3,80){\bfseries (b)}
    \end{overpic}    
    \hfill    
    \begin{overpic}[width=0.32\textwidth]{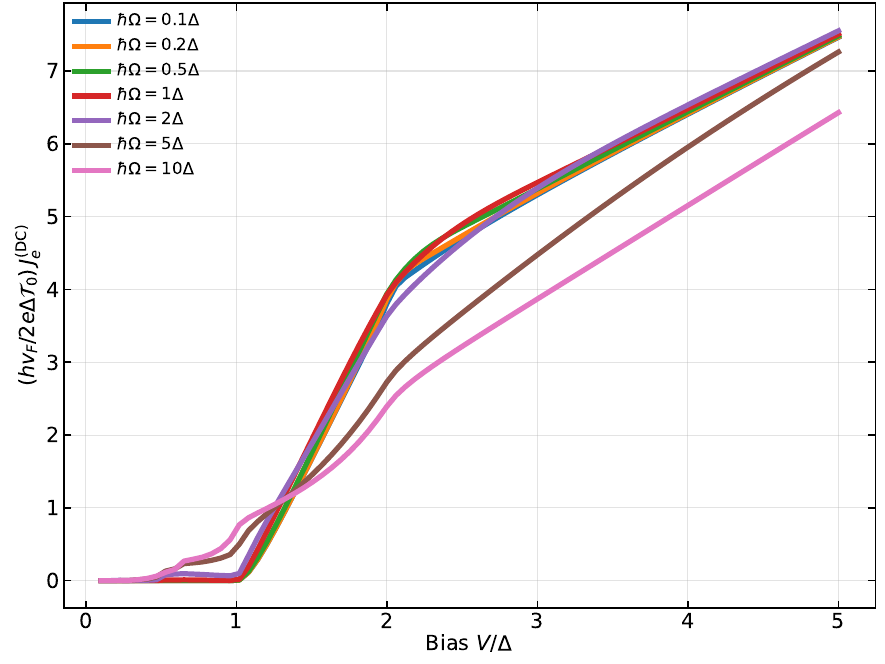}
        \put(3,80){\bfseries (c)}
    \end{overpic}

    \par
    \vspace{20pt}

    \begin{overpic}[width=0.24\textwidth]{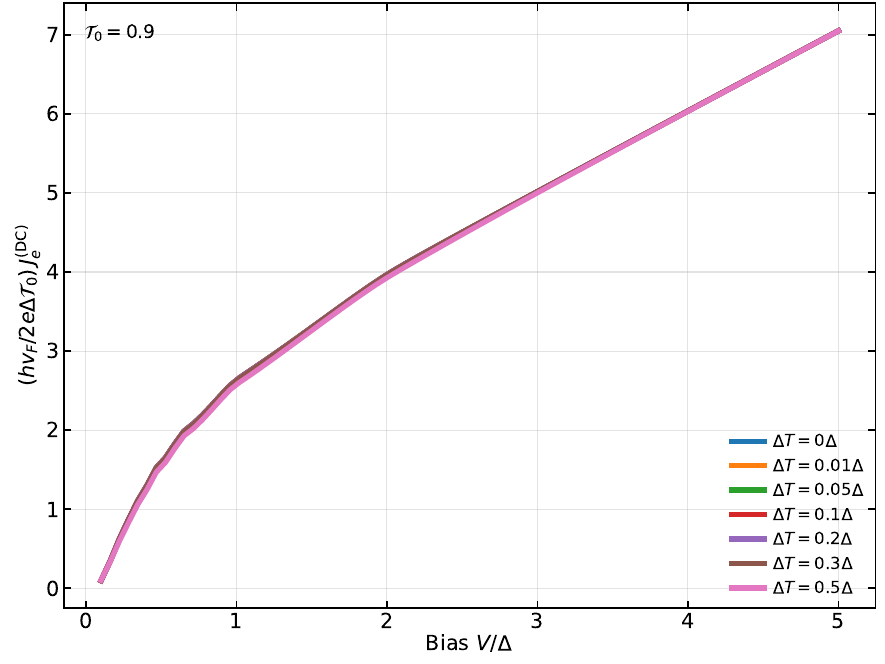}
        \put(3,80){\bfseries (d)}
    \end{overpic}    
    \hfill    
    \begin{overpic}[width=0.24\textwidth]{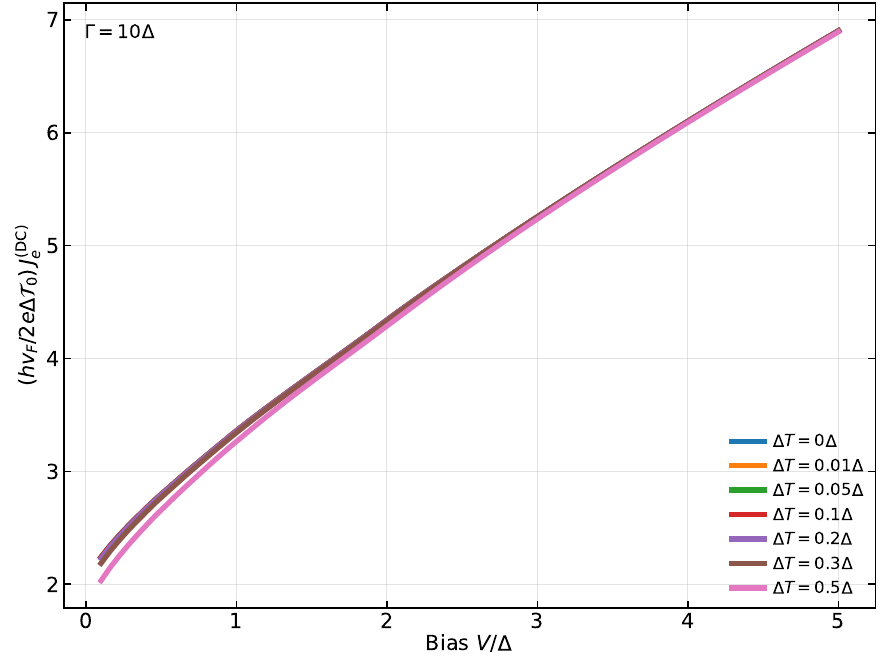}
        \put(3,80){\bfseries (e)}
    \end{overpic}    
    \hfill    
    \begin{overpic}[width=0.24\textwidth]{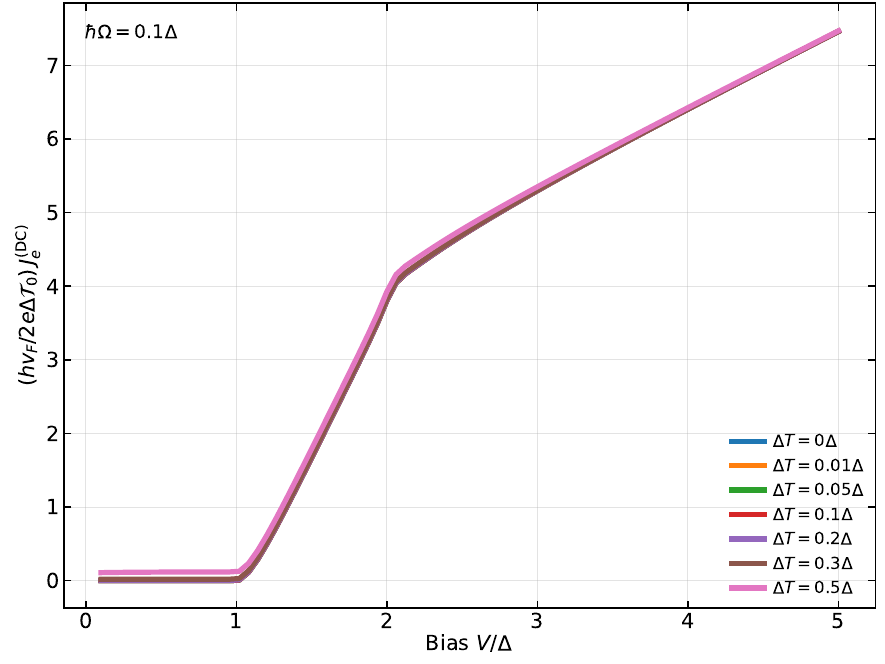}
        \put(3,80){\bfseries (f)}
    \end{overpic}    
    \hfill    
    \begin{overpic}[width=0.24\textwidth]{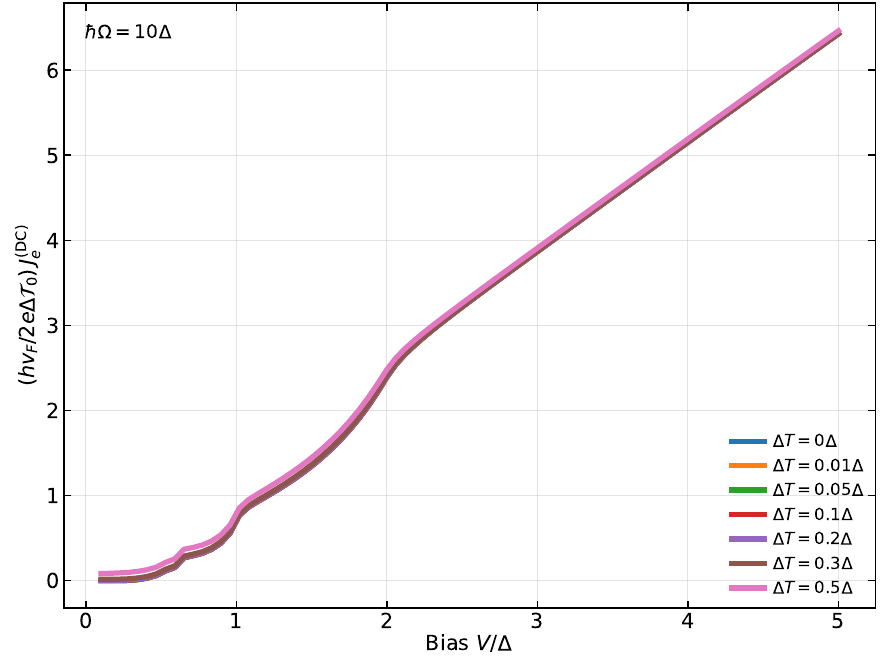}
        \put(3,80){\bfseries (g)}
    \end{overpic}

    \caption{
        Bias dependence of the dc electric current.
        Panels~(a)-(c) use 
        the energy-independent $\mathcal{T}_{\mathrm{const}}$,
        the quantum-dot-type $\mathcal{T}_{\mathrm{Qdot}}$, and
        the QPC-type $\mathcal{T}_{\mathrm{QPC}}$ profiles, respectively, at $\Delta T=0$.
        Panels (d)-(g) compare  several temperature differences 
        for (d) the energy-independent profile with $\cT_0 = 0.9$,
        (e) the quantum-dot-type profile with $\Gamma = 10\Delta$,
        (f) the QPC profile with $\hbar\Omega = 0.1\Delta$ and
        (g) $\hbar\Omega = 10\Delta$.
        Panel (a) reproduces the result of Averin and Bardas~\cite{AB}, and panel (b) the resonant-level result of Levy Yeyati \textit{et al.}~\cite{Levy}.
        The legends give the parameter value.
    }
    \label{fig:electric-current_bias}
\end{figure*}

\begin{figure*}[t]
    \centering

    \begin{overpic}[width=0.32\textwidth]{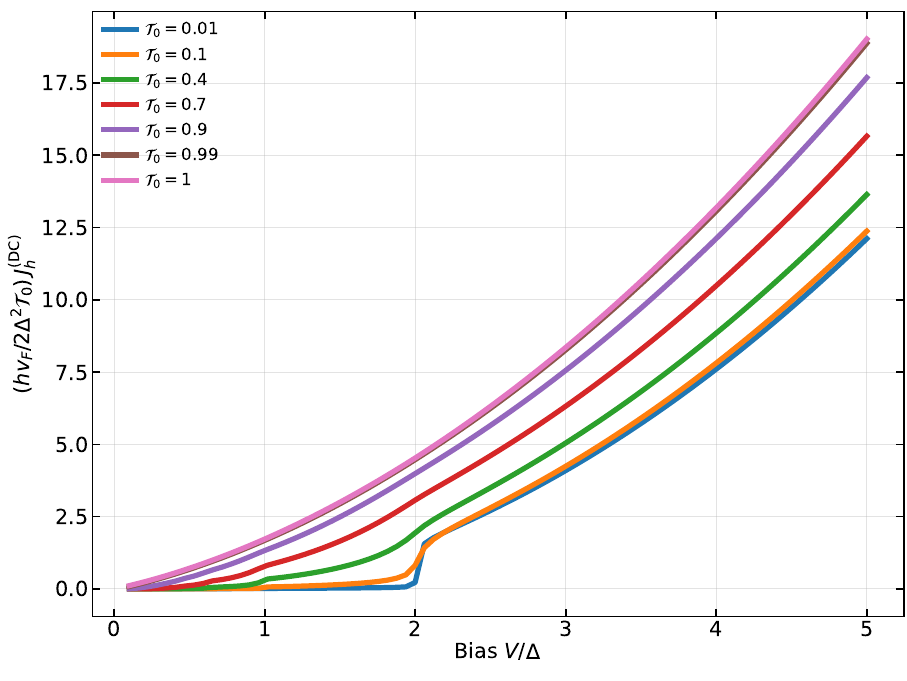}
        \put(3,80){\bfseries (a)}
    \end{overpic}    
    \hfill    
    \begin{overpic}[width=0.32\textwidth]{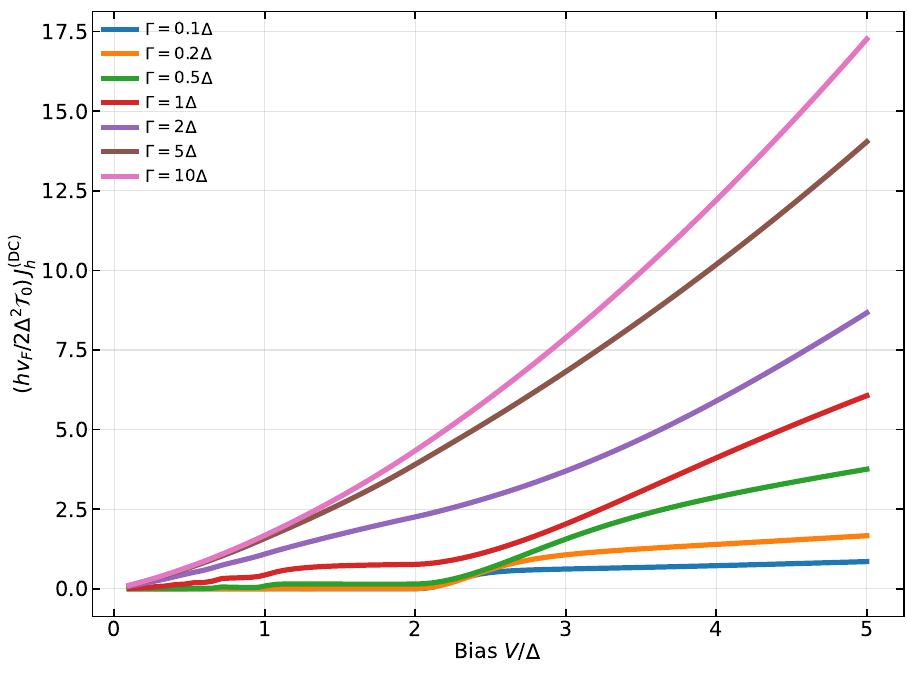}
        \put(3,80){\bfseries (b)}
    \end{overpic}    
    \hfill    
    \begin{overpic}[width=0.32\textwidth]{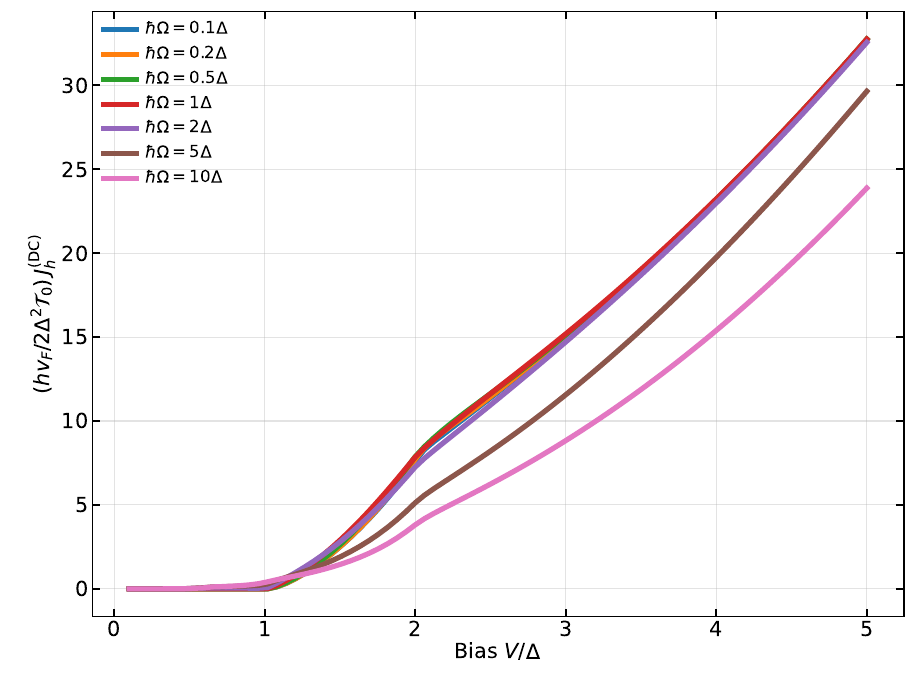}
        \put(3,80){\bfseries (c)}
    \end{overpic}

    \par
    \vspace{20pt}
    
    \begin{overpic}[width=0.24\textwidth]{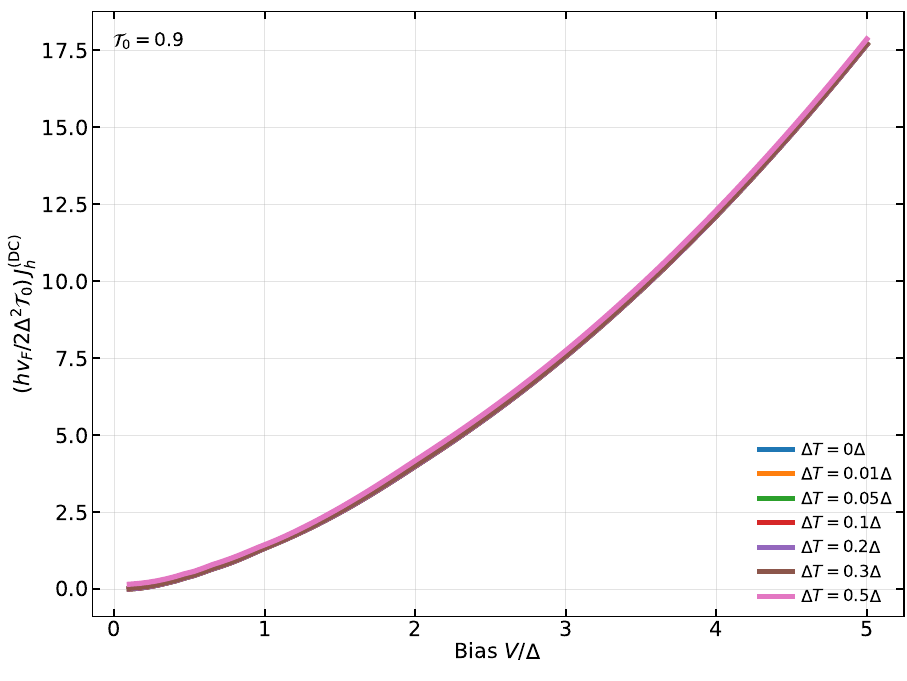}
        \put(3,80){\bfseries (d)}
    \end{overpic}    
    \hfill    
    \begin{overpic}[width=0.24\textwidth]{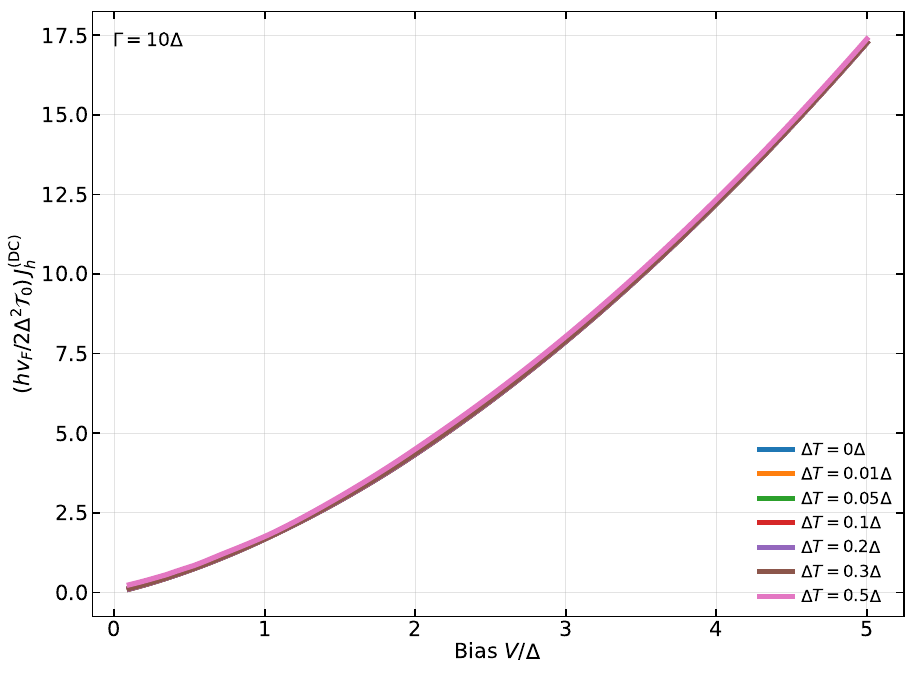}
        \put(3,80){\bfseries (e)}
    \end{overpic}    
    \hfill    
    \begin{overpic}[width=0.24\textwidth]{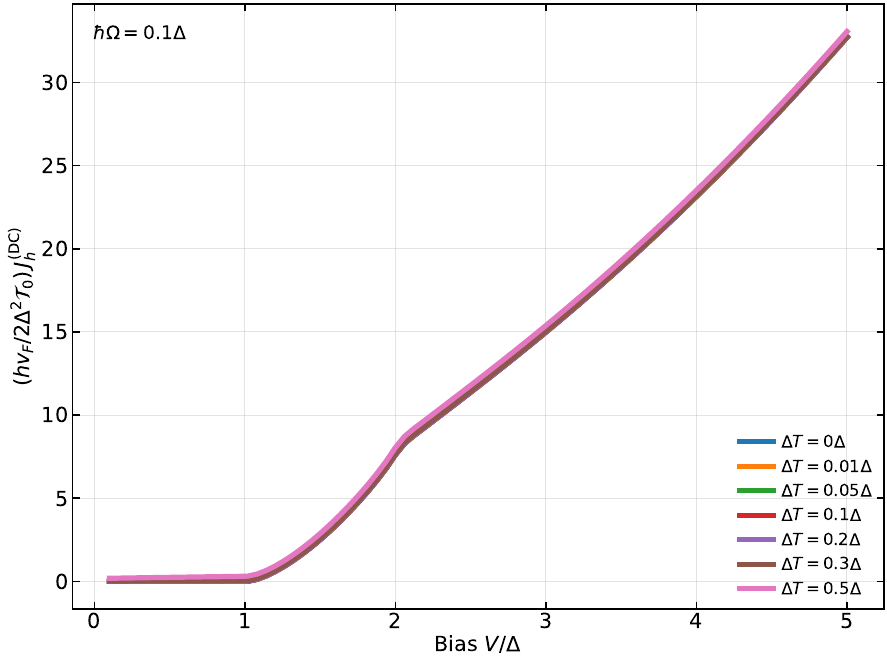}
        \put(3,80){\bfseries (f)}
    \end{overpic}
    \hfill    
    \begin{overpic}[width=0.24\textwidth]{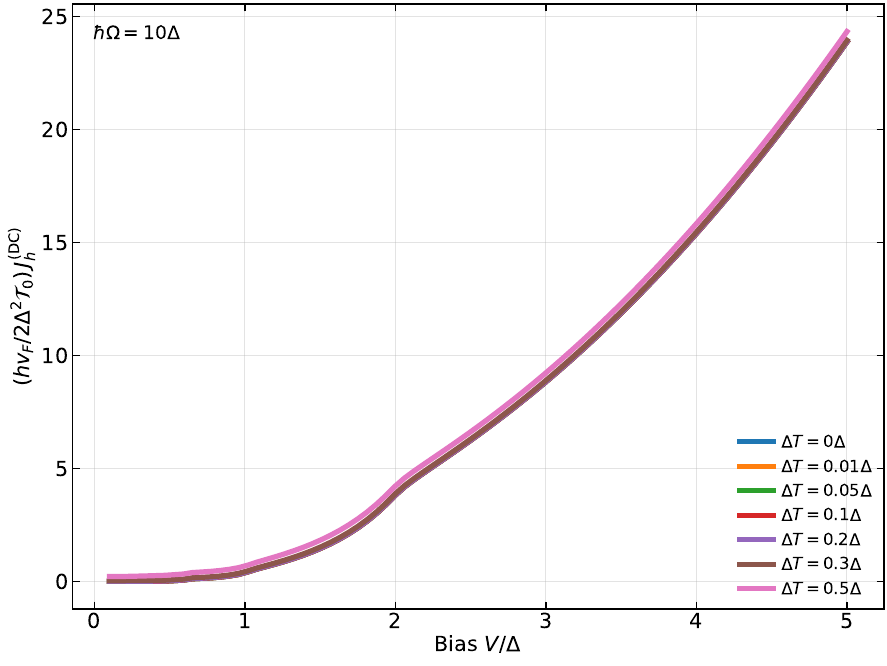}
        \put(3,80){\bfseries (g)}
    \end{overpic}

    \caption{
        Bias dependence of the dc heat current.
        Panels~(a)-(c) use 
        the energy-independent $\mathcal{T}_{\mathrm{const}}$,
        (b) the quantum-dot-type $\mathcal{T}_{\mathrm{Qdot}}$, and
        (c) the QPC $\mathcal{T}_{\mathrm{QPC}}$ profiles, respectively at $\Delta T=0$.
        Panels (d)-(g) compare several temperature differences for 
        (d) the energy-independent profile with $\cT_0 = 0.9$,
        (e) the quantum-dot-type profile with $\Gamma = 10\Delta$,
        (f) the QPC profile with $\hbar\Omega = 0.1\Delta$ and
        (g)  $\hbar\Omega = 10\Delta$.
        The legends give the parameter values.
    }
    \label{fig:heat-current_bias}
\end{figure*}

Figures~\ref{fig:electric-current_bias} 
and~\ref{fig:heat-current_bias} show the bias dependence
of the dc electric and heat current, respectively.
Panels~(a)-(c) compare the three transmission profiles
at $\Delta T=0$, while panels (d)-(g) show the effect
of a temperature difference for selected transmission parameters.

At low bias, the electric current exhibits
nonlinear features associated with the superconducting
gap and MAR. At the upper end of the plotted bias range,
the current becomes approximately linear in $V$ for
the energy-independent and QPC-like profiles.
The quantum-dot-like profile shows a stronger dependence
on the resonance width: broad resonances yield a more nearly linear current-bias characteristic, whereas
narrow resonances produce pronounced nonmonotonic features and a weak increase, or even a decrease,
of the current at higher bias.
Thus, the approach to an approximately linear current-bias characteristic is not uniform across the transmission profiles considered.

The heat current shows a distinct bias dependence.
Its magnitude is small at low bias and increases
as the bias is raised, yet substantial curvature
remains over the plotted range.
In particular, the energy-independent and QPC profiles,
as well as broad quantum-dot-like resonances, exhibit
a pronounced upward curvature. These results
therefore do not support a general high-bias proportionality between $\jh$ and $V$.
The finite-bias heat-current response should be distinguished from the linear thermoelectric cross response described by $M$ in Eq.~\eqref{eq:thermoelectric_transport}, which is defined near equilibrium.

For the parameter sets shown in Figs.~\ref{fig:electric-current_bias}(d)--(g) and~\ref{fig:heat-current_bias}(d)--(g), changing the temperature difference produces
relatively small modifications of the bias-dependent
curves on the plotted current scales.
The overall variation in these panels is dominated by
the applied bias.

\subsection{Temperature-difference dependence}

\begin{figure*}[t]
    \centering

    \begin{overpic}[width=0.32\textwidth]{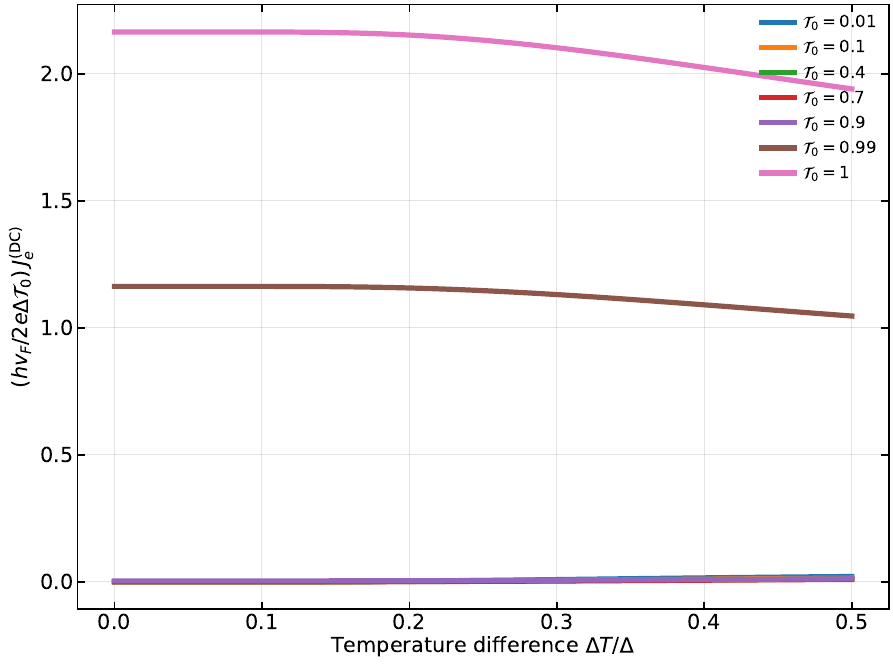}
        \put(3,80){\bfseries (a)}
    \end{overpic}
    \hfill
    \begin{overpic}[width=0.32\textwidth]{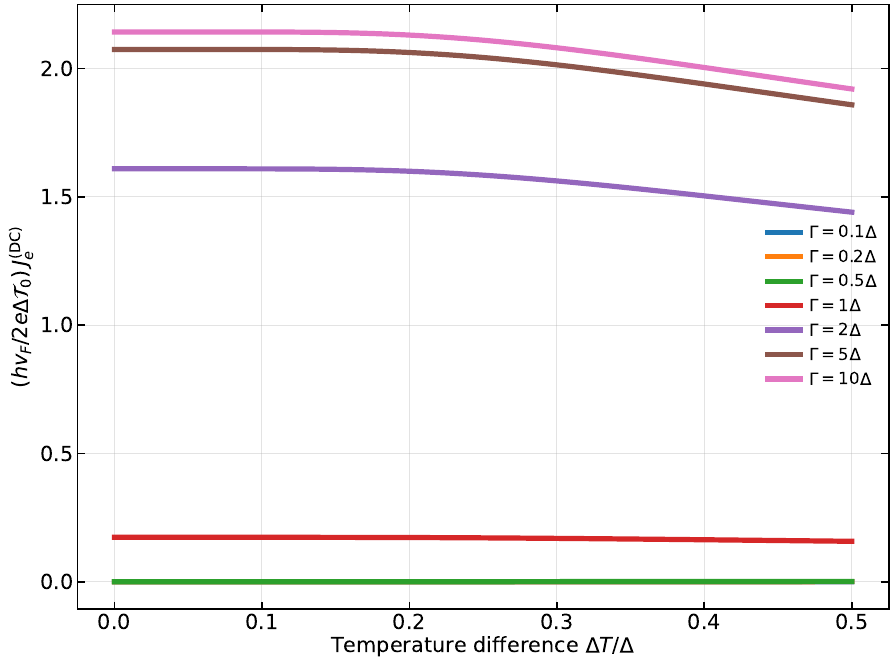}
        \put(3,80){\bfseries (b)}
    \end{overpic}
    \hfill
    \begin{overpic}[width=0.32\textwidth]{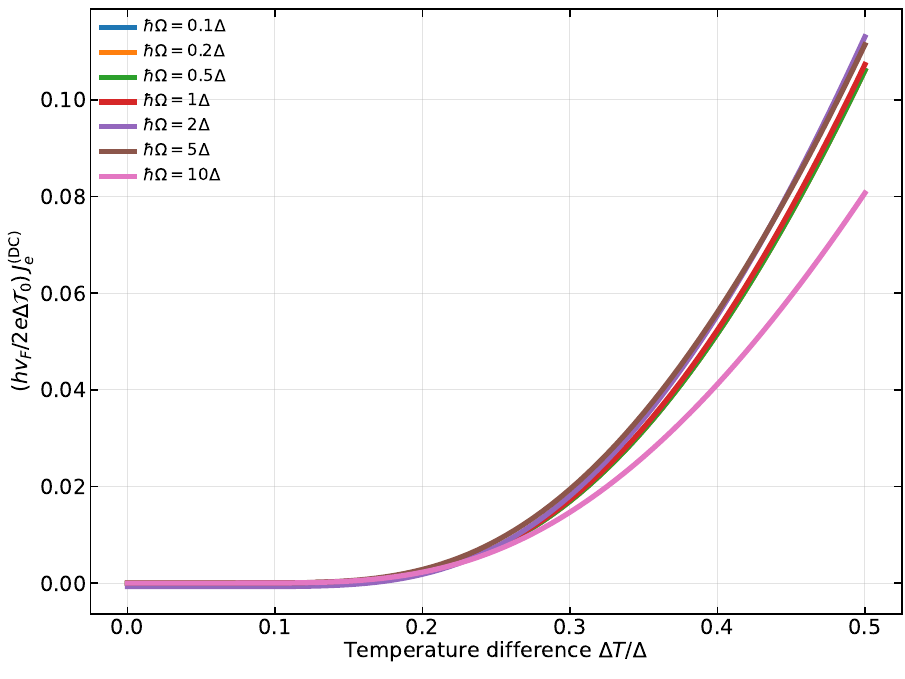}
        \put(3,80){\bfseries (c)}
    \end{overpic}
    
    \par
    \vspace{20pt}

    \begin{overpic}[width=0.24\textwidth]{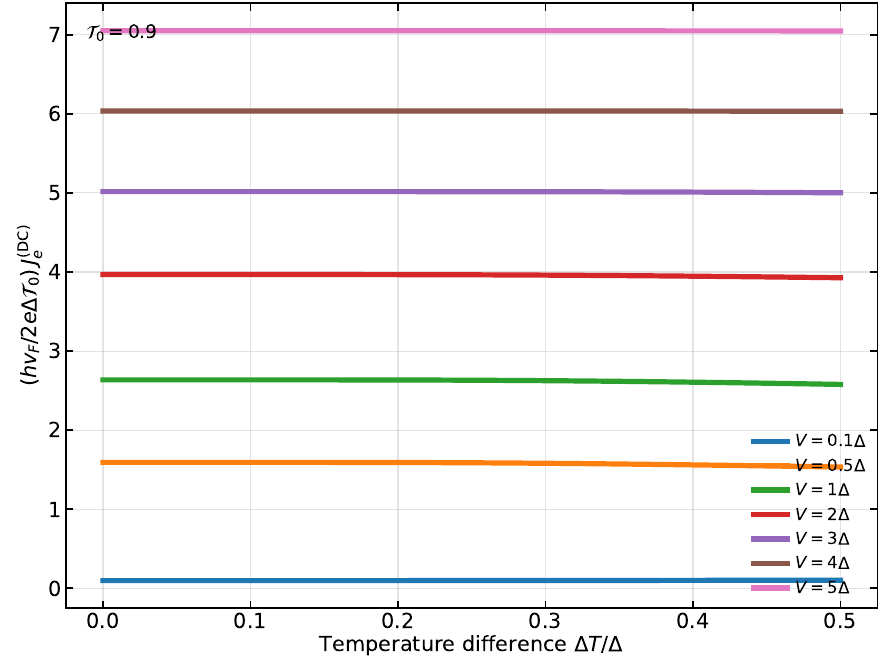}
        \put(3,80){\bfseries (d)}
    \end{overpic}
    \hfill
    \begin{overpic}[width=0.24\textwidth]{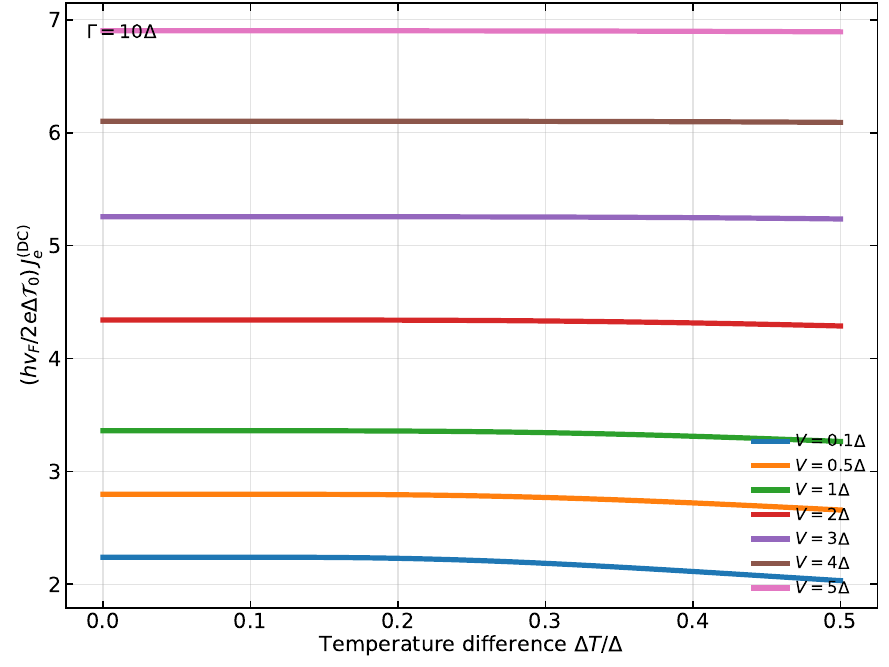}
        \put(3,80){\bfseries (e)}
    \end{overpic}
    \hfill
    \begin{overpic}[width=0.24\textwidth]{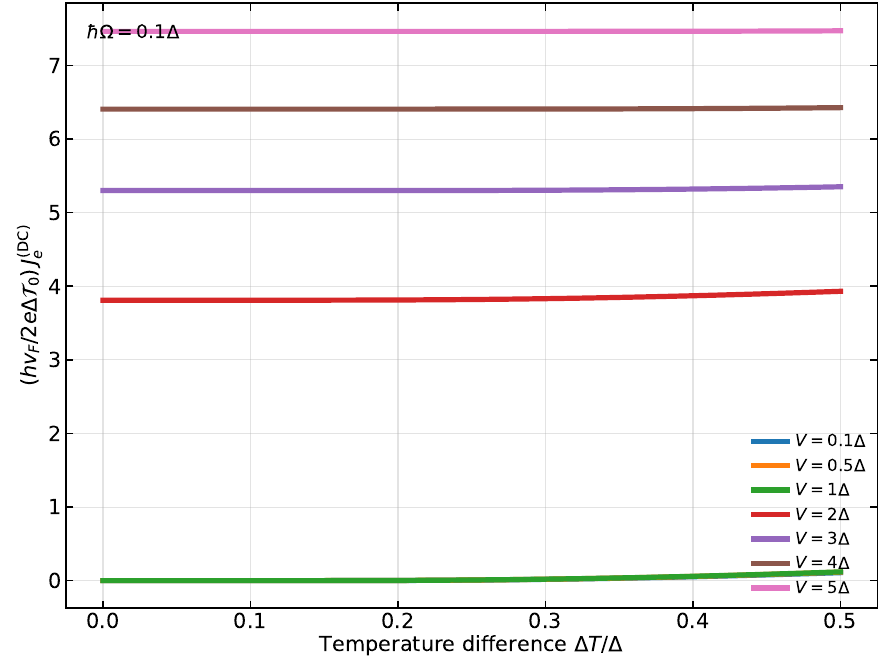}
        \put(3,80){\bfseries (f)}
    \end{overpic}
    \hfill
    \begin{overpic}[width=0.24\textwidth]{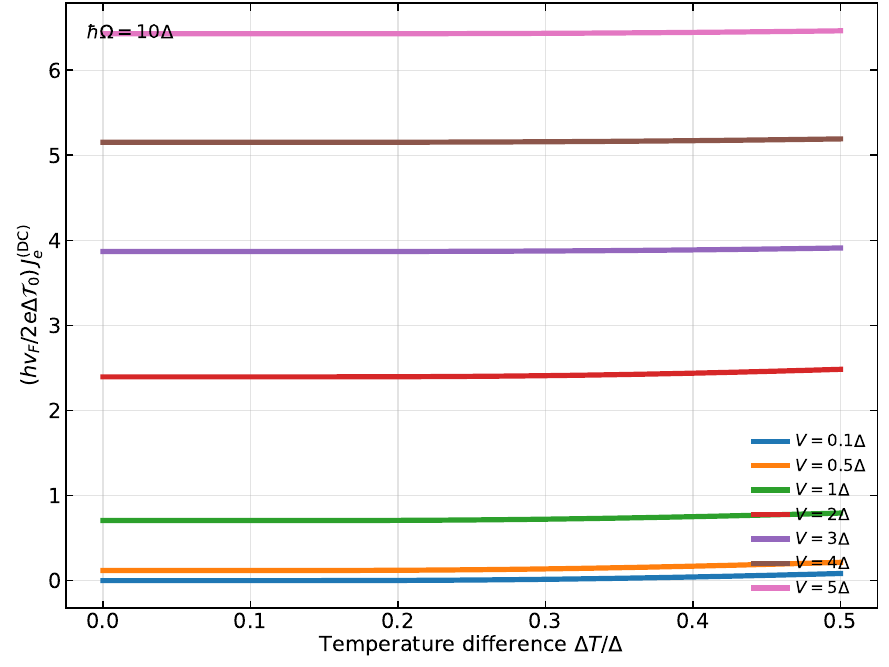}
        \put(3,80){\bfseries (g)}
    \end{overpic}
    
    \caption{
        Temperature-difference dependence of the dc electric current. Panels~(a)-(c) use 
        the energy-independent $\mathcal{T}_{\mathrm{const}}$,
        (b) the quantum-dot-type $\mathcal{T}_{\mathrm{Qdot}}$, and
        (c) the QPC $\mathcal{T}_{\mathrm{QPC}}$ profiles, respectively, at $V=0.05\Delta$.
        Panels (d)-(g) compare several biases for
        (d) the energy-independent profile with $\cT_0 = 0.9$,
        (e) the quantum-dot-type profile with $\Gamma = 10\Delta$,
        (f) the QPC profile with $\hbar\Omega = 0.1\Delta$ and
        (g) $\hbar\Omega = 10\Delta$.
        The legends give the parameter values.
    }
    \label{fig:electric-current_dT}
\end{figure*}

\begin{figure*}[t]
    \centering

    \begin{overpic}[width=0.32\textwidth]{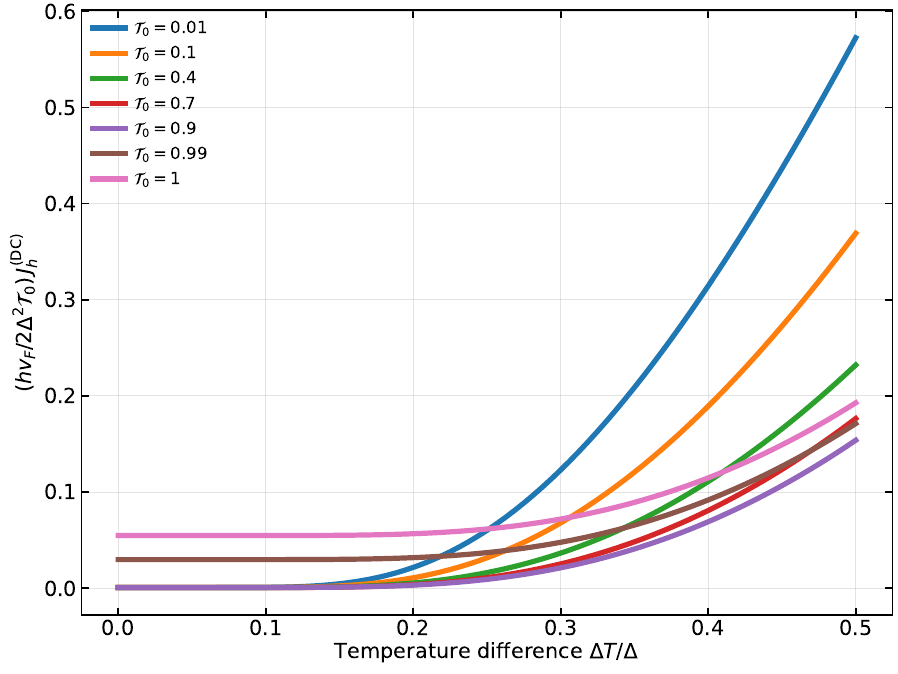}
        \put(3,80){\bfseries (a)}
    \end{overpic}
    \hfill
    \begin{overpic}[width=0.32\textwidth]{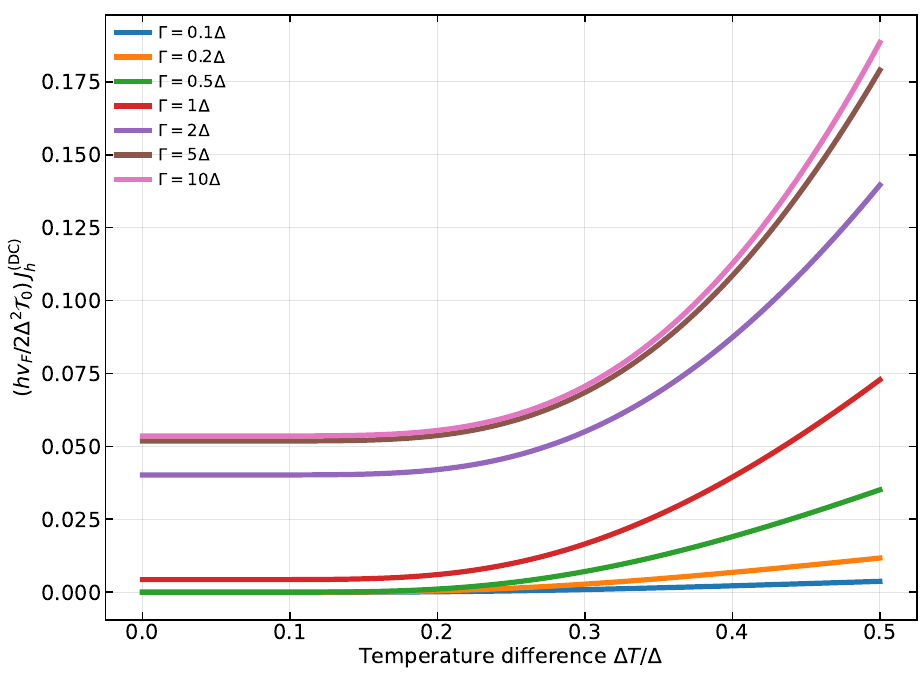}
        \put(3,80){\bfseries (b)}
    \end{overpic}
    \hfill
    \begin{overpic}[width=0.32\textwidth]{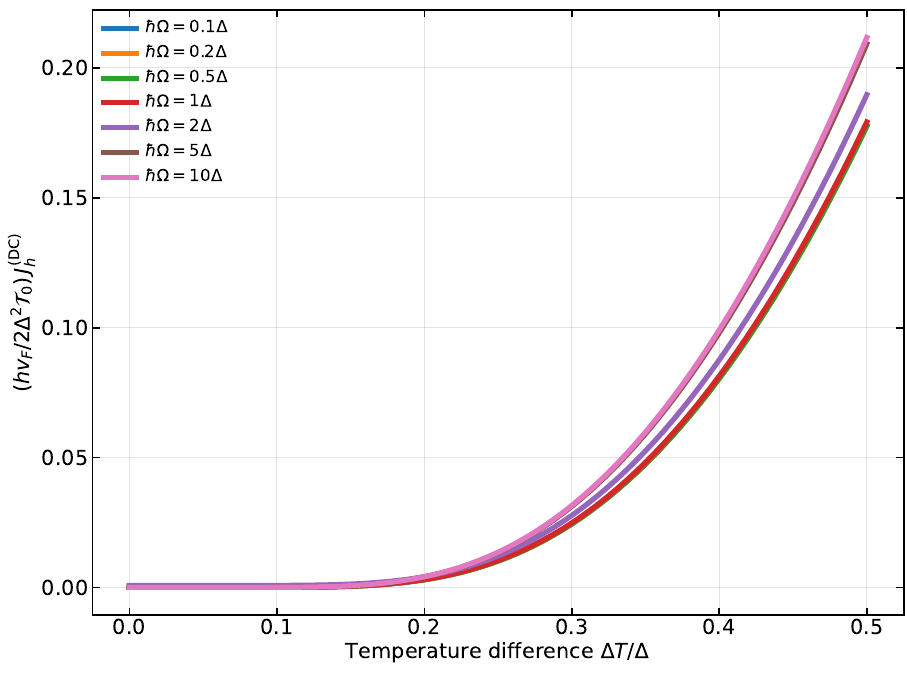}
        \put(3,80){\bfseries (c)}
    \end{overpic}

    \par
    \vspace{20pt}

    \begin{overpic}[width=0.24\textwidth]{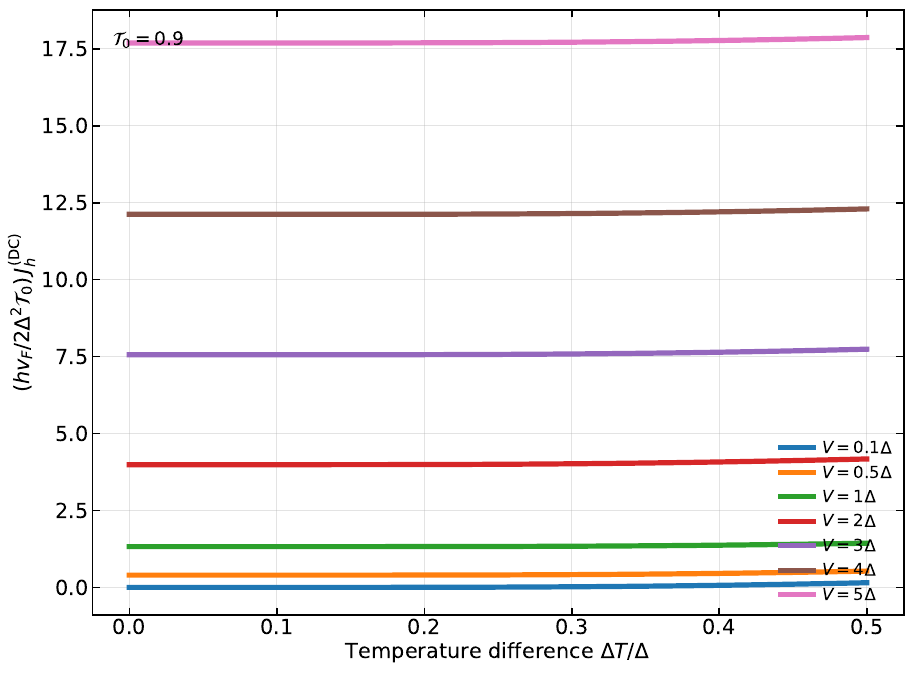}
        \put(3,80){\bfseries (d)}
    \end{overpic}
    \hfill
    \begin{overpic}[width=0.24\textwidth]{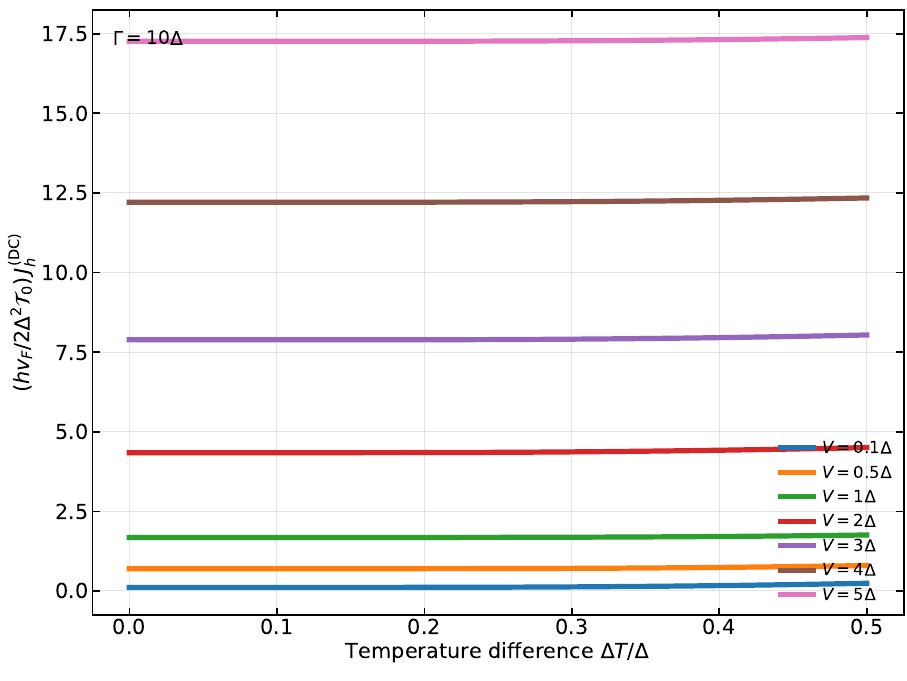}
        \put(3,80){\bfseries (e)}
    \end{overpic}
    \hfill
    \begin{overpic}[width=0.24\textwidth]{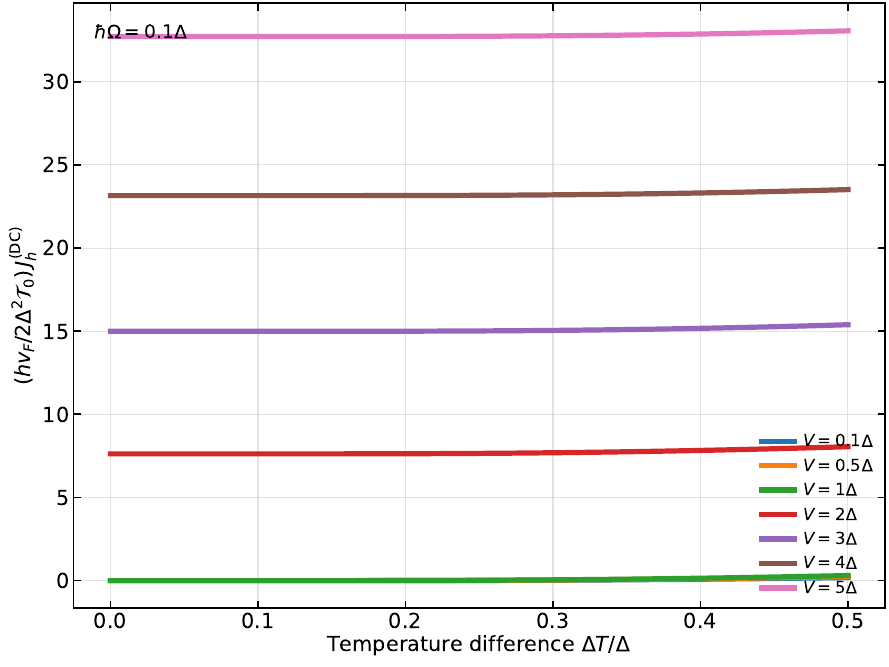}
        \put(3,80){\bfseries (f)}
    \end{overpic}
    \hfill
    \begin{overpic}[width=0.24\textwidth]{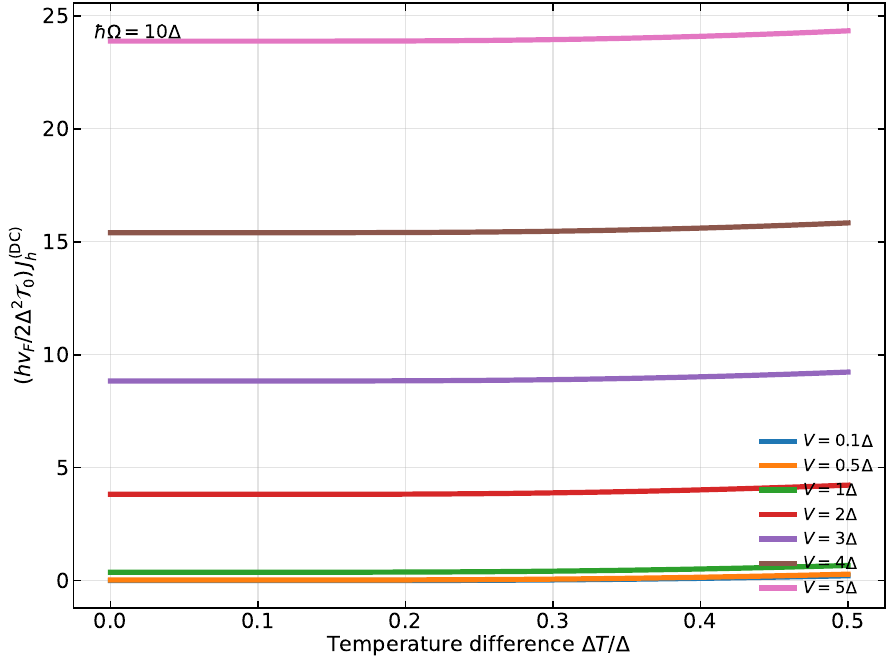}
        \put(3,80){\bfseries (g)}
    \end{overpic}

    \caption{
        Temperature-difference dependence of the dc heat current. 
        Panels~(a)-(c) use 
       the energy-independent $\mathcal{T}_{\mathrm{const}}$,
        (b) the quantum-dot-type $\mathcal{T}_{\mathrm{Qdot}}$, and
        (c) the QPC $\mathcal{T}_{\mathrm{QPC}}$ profiles, respectively at $V=0.05\Delta$.
        Panels (d)-(g) compare several biases for
        (d) the energy-independent profile with $\cT_0 = 0.9$,
        (e) the quantum-dot-type profile with $\Gamma = 10\Delta$,
        (f) the QPC profile with $\hbar\Omega = 0.1\Delta$ and
        (g)  $\hbar\Omega = 10\Delta$.
        The legends give the parameter values.
    }
    \label{fig:heat-current_dT}
\end{figure*}

Figures~\ref{fig:electric-current_dT} and~\ref{fig:heat-current_dT} show the temperature-difference dependence of the dc electric and heat currents, respectively.
In panels~(a)-(c), the bias is fixed at the small but
finite value $V=0.05\Delta$.

For the energy-independent and quantum-dot-like profiles,
the electric current depends only weakly on $\Delta T$,
as shown in Figs.~\ref{fig:electric-current_dT}~(a) 
and~\ref{fig:electric-current_dT}~(b).
This dependence is not entirely absent: for large ${\cal T}_0$ and broad resonances, the current decreases modestly as the temperature difference increases.
More generally, the magnitude and sign of the smaller
temperature-induced changes depend on the transmission parameters.

In contrast, the QPC profile produces a 
pronounced nonlinear increase in the electric current
with $\Delta T$, as shown in Fig.~\ref{fig:electric-current_dT}. The contrast between the three profiles
demonstrates that the temperature dependence of the 
electric current is sensitive to the energy dependence of
the normal-region transmission. Since these calculations are performed at finite bias, they characterize the thermal modulation of a bias-driven current rather than directly determining the equilibrium thermoelectric coefficient $L$.

The heat current exhibits a nonlinear dependence on $\Delta T$ for all three transmission profiles,
as shown in Figs.~\ref{fig:heat-current_dT}~(a)-(c).
Both the magnitude of the current and its temperature
dependence vary with the transmission parameters.
Over the full temperature-difference range investigated,
the response cannot be represented by a single linear term proportional to $K\Delta T$.

Figures~\ref{fig:electric-current_dT}~(d)-(g)
and~\ref{fig:heat-current_dT}~(d)-(g)
compare the temperature-difference dependence
at several biases. Changing $V$ substantially shifts
the current levels, while the variation with $\Delta T$
at each fixed bias is comparatively modest on the plotted
scales. These panels therefore show a strong bias dependence of the current background, rather than
an overall insensitivity to the bias.

Finally, finite electric and heat currents remain
at $\Delta T=0$ for some of the parameter sets in
Figs.~\ref{fig:electric-current_dT}~(a)-(c)
and~\ref{fig:heat-current_dT}~(a)-(c).
These are bias-driven contributions arising 
from the nonzero bias used in the calculations and 
should not be interpreted as temperature-driven currents.
The results in these panels should therefore be understood as temperature responses at a small but finite
bias, rather than as strictly zero-bias transport characteristics.

\clearpage
\section{Conclusion}
\label{sec:conclusion}

We investigated the phase-independent dc electric and heat currents through a one-dimensional superconductor-normal-metal-superconductor junction driven by chemical-potential
 and temperature differences. 
The scattering framework incorporates MARs and retains
the energy dependence of normal-region transmission,
allowing us to compare energy-independent, quantum-dot-type,
and QPC profiles.

At low bias, the electric and heat currents exhibit 
nonlinear features associated with the superconducting
gap and MARs. At higher bias,
the electric current becomes approximately linear for
the energy-independent and quantum-point-contact profiles
over the range studied, whereas the heat current retains
appreciable nonlinearity. At a small finite bias,
the electric current varies only weakly with the temperature difference for the energy-independent and quantum-dot-type profiles, but exhibits a pronounced nonlinear temperature dependence for the QPC profile. The heat current depends nonlinearly on
the temperature difference for all three profiles over the parameter range investigated.
These differences highlight the importance of 
examining both currents when characterizing transport
beyond linear response.

Superconducting junctions can also exhibit phase-dependent
thermoelectric and thermal responses,
including contributions to the linear-response coefficients
$L$ and $K$~\cite{Maki,Zhao2003,Zhao2004,Pershoguba,Giazotto}.
In the cold-atom experiments discussed above,
particle and entropy currents are inferred from repeated
destructive measurements of the reservoirs at 
different evolution times~\cite{Krinner}.
If the relative phase is uncontrolled and varies
between experimental realizations, the measured currents
reflect an average over its distribution. Oscillatory contributions from the ac Josephson effect can also be suppressed by averaging over
times much longer than the Josephson period~\cite{Husmann}.
Quantitative comparison therefore requires
appropriate phase and time averaging consistent with the experimental
preparation and temporal resolution.

Furthermore, experiments at and near unitarity probe a strongly
interacting regime in which the pairing gap is not necessarily small
compared with the chemical potential. The validity of the Andreev
approximation adopted here is therefore not guaranteed 
in this regime~\cite{Hofmann}.
A unified treatment of nonlinear particle (electric) and heat transport that incorporates phase dynamics, the experimental averaging
procedure, and effects beyond the Andreev approximation 
remains a task for future work.

\section*{acknowledgments}
SU thanks D. Bertolusso, T. Bonaccorsi, T. Esslinger, T. Giamarchi, M. M\"{u}hlematter, M. Talebi, S. Wili, and Y. Yudkin for discussions.
SU is supported by JST PRESTO (JPMJPR2351) and JSPS
KAKENHI (JP25K07191).

\appendix
\onecolumngrid 
\section{Derivation of the recurrence relation}
\label{app:derivation_recurrence}

In this Appendix, we derive Eqs.~\eqref{eq:recurrence-formula} and \eqref{eq:beta-bp} from Eq.~\eqref{eq:scattering-relation_compact}.

\subsection{Elimination of the scattering amplitudes}
\label{subsec:elimination_of_the_scattering_amplitudes}

We begin with Eq.~\eqref{eq:scattering-relation_compact}, which reads
\begin{equation}
    \beta_j^{(p)}(n)
    =
    \delta_{j, k} \delta_{p, q} \delta_{n, 0}
    I_j^{(p)}(n)
    +
    A_j^{(\bp)}(n)
    \beta_j^{(\bp)}(n)
    +
    \delta_{\bj, k} \delta_{p, q} \delta_{n \pm 1, 0}
    J_{\bj}^{(p)}(n \pm 1)
    +
    B_{\bj}^{(\bp)}(n \pm 1)
    \beta_{\bj}^{(\bp)}(n \pm 1).
\label{eq:app_1}
\end{equation}
Using the notation rules in Eq.~\eqref{eq:sideband_index_rules}, we replace $p\to\bp$ in Eq.~\eqref{eq:app_1}, which gives
\begin{equation}
    \beta_j^{(\bp)}(n)
    =
    \delta_{j, k} \delta_{\bp, q} \delta_{n,0}
    I_j^{(\bp)}(n)
    +
    A_j^{(p)}(n)
    \beta_j^{(p)}(n)
    +
    \delta_{\bj, k} \delta_{\bp, q} \delta_{n \mp 1, 0}
    J_{\bj}^{(\bp)}(n \mp 1)
    +
    B_{\bj}^{(p)}(n \mp 1)
    \beta_{\bj}^{(p)}(n \mp 1).
\label{eq:app_2}
\end{equation}
Replacing $j\to\bj$ in Eq.~\eqref{eq:app_2} then yields
\begin{equation}
    \beta_{\bj}^{(\bp)}(n)
    =
    \delta_{\bj, k} \delta_{\bp, q} \delta_{n,0}
    I_{\bj}^{(p)}(n)
    +
    A_{\bj}^{(\bp)}(n)
    \beta_{\bj}^{(p)}(n)
    +
    \delta_{j, k} \delta_{\bp, q} \delta_{n \pm 1, 0}
    J_{j}^{(\bp)}(n \pm 1)
    +
    B_{j}^{(p)}(n \pm 1)
    \beta_{j}^{(p)}(n \pm 1).
\label{eq:app_3}
\end{equation}
Shifting the sideband index as $n\to n\mp1$ in Eq.~\eqref{eq:app_3} and solving for $ \beta_{\bj}^{(p)}(n \mp 1)$, we find
\begin{equation}
    \beta_{\bj}^{(p)}(n \mp 1)
    =
    \frac{1}{
        A_{\bj}^{(p)}(n \mp 1)
    }
    \Bigg[
        \beta_{\bj}^{(\bp)}(n \mp 1)
        -
        B_j^{(p)}(n)
        \beta_j^{(p)}(n)
        -
        \delta_{\bj, k} \delta_{\bp, q} \delta_{n \mp 1, 0}
        I_{\bj}^{(\bp)}(n \mp 1)
        -
        \delta_{j, k} \delta_{\bp, q} \delta_{n, 0}
        J_j^{(\bp)}(n)
    \Bigg].
\label{eq:app_4}
\end{equation}
Substituting Eq.~\eqref{eq:app_4} into Eq.~\eqref{eq:app_2} and rearranging, we  obtain
\begin{align}
    &
        \begin{aligned}
                \beta_{\bj}^{(\bp)}(n \mp 1)
                ={}
            &
                \left(
                    B_j^{(p)}(n)
                    -
                    \frac{
                        A_{\bj}^{(p)}(n \mp 1)
                    }{
                        B_{\bj}^{(p)}(n \mp 1)
                    }
                    A_j^{(p)}(n)
                \right)
                \beta_j^{(p)}(n)
                +
                \frac{
                    A_{\bj}^{(p)}(n \mp 1)
                }{
                    B_{\bj}^{(p)}(n \mp 1)
                }
                \beta_j^{(\bp)}(n)
            \\ &
                +
                \delta_{j, k} \delta_{\bp, q} \delta_{n, 0}
                \left(
                    J_j^{(\bp)}(n)
                    -
                    \frac{
                        A_{\bj}^{(p)}(n \mp 1)
                    }{
                        B_{\bj}^{(p)}(n \mp 1)
                    }
                    I_j^{(\bp)}(n)
                \right)
            \\ &
                +
                \delta_{\bj, k} \delta_{\bp q} \delta_{n \mp 1, 0}
                \left(
                    I_{\bj}^{(\bp)}(n \mp 1)
                    -
                    \frac{
                        A_{\bj}^{(p)}(n \mp 1)
                    }{
                        B_{\bj}^{(p)}(n \mp 1)
                    }
                    J_{\bj}^{(\bp)}(n \mp 1)
                \right)
        \end{aligned}
    \\ &
        =
        C_j^{(p)}(n) \beta_j^{(p)}(n)
        +
        D_j^{(\bp)}(n) \beta_j^{(\bp)}(n)
        +
        \delta_{j, k} \delta_{\bp, q} \delta_{n, 0}
        K_j^{(\bp)}(n)
        +
        \delta_{\bj, k} \delta_{\bp q} \delta_{n \mp 1, 0}
        L_{\bj}^{(\bp)}(n \mp 1).
\label{eq:app_5}
\end{align}
Applying the shift $n\to n\pm2$ to Eq.~\eqref{eq:app_5} gives
\begin{equation}
    \beta_{\bj}^{(\bp)}(n \pm 1)
    =
    C_j^{(p)}(n \pm 2) \beta_j^{(p)}(n \pm 2)
    +
    D_j^{(\bp)}(n \pm 2) \beta_j^{(\bp)}(n \pm 2)
    +
    \delta_{j, k} \delta_{\bp, q} \delta_{n \pm 2, 0}
    K_j^{(\bp)}(n \pm 2)
    +
    \delta_{\bj, k} \delta_{\bp q} \delta_{n \pm 1, 0}
    L_{\bj}^{(\bp)}(n \pm 1).
\label{eq:app_6}
\end{equation}
Substituting Eq.~\eqref{eq:app_6} into Eq.~\eqref{eq:app_1} yields
\begin{align}
    &
        \begin{aligned}
                \beta_j^{(p)}(n)
                ={}
            &
                B_{\bj}^{(\bp)}(n \pm 1)
                C_j^{(p)}(n \pm 2)
                \beta_j^{(p)}(n \pm 2)
                +
                A_j^{(\bp)}(n)
                \beta_j^{(\bp)}(n)
                +
                B_{\bj}^{(\bp)}(n \pm 1)
                D_j^{(\bp)}(n \pm 2)
                \beta_j^{(\bp)}(n \pm 2)
            \\ &
                +
                \delta_{j, k} \delta_{p, q} \delta_{n, 0}
                I_j^{(p)}(n)
                +
                \delta_{\bj, k} \delta_{n \pm 1, 0} \delta_{p, q}
                J_{\bj}^{(p)}(n \pm 1)
                +
                \delta_{\bj, k} \delta_{n \pm 1, 0} \delta_{\bp, q}
                B_{\bj}^{(\bp)}(n \pm 1)
                L_{\bj}^{(\bp)}(n \pm 1)
            \\ &
                +
                \delta_{j, k} \delta_{\bp, q} \delta_{n \pm 2, 0}
                B_{\bj}^{(\bp)}(n \pm 1)
                K_j^{(\bp)}(n \pm 2)
        \end{aligned}
    \\ &
        \begin{aligned}
                =
            &
                E_j^{(p)}(n \pm 2)
                \beta_j^{(p)}(n \pm 2)
                +
                A_j^{(\bp)}(n)
                \beta_j^{(\bp)}(n)
                +
                F_j^{(\bp)}(n \pm 2)
                \beta_j^{(\bp)}(n \pm 2)
                +
                \delta_{j, k} \delta_{p, q} \delta_{n, 0}
                I_j^{(p)}(n)
            \\ &
                +
                \delta_{\bj, k} \delta_{p, q} \delta_{n \pm 1, 0}
                J_{\bj}^{(p)}(n \pm 1)
                +
                \delta_{\bj, k} \delta_{\bp, q} \delta_{n \pm 1, 0}
                M_{\bj}^{(\bp)}(n \pm 1)
                +
                \delta_{j, k} \delta_{\bp, q} \delta_{n \pm 2, 0}
                N_j^{(\bp)}(n \pm 2).
        \end{aligned}
\label{eq:app_7}
\end{align}
Replacing $p\to\bp$ in Eq.~\eqref{eq:app_7},
we obtain
\begin{equation}
\begin{aligned}
        \beta_j^{(\bp)}(n)
        ={}
    &
        E_j^{(\bp)}(n \mp 2)
        \beta_j^{(\bp)}(n \mp 2)
        +
        A_j^{(p)}(n)
        \beta_j^{(p)}(n)
        +
        F_j^{(p)}(n \mp 2)
        \beta_j^{(p)}(n \mp 2)
        +
        \delta_{j, k} \delta_{\bp, q} \delta_{n, 0}
        I_j^{(\bp)}(n)
    \\ &
        +
        \delta_{\bj, k} \delta_{\bp, q} \delta_{n \mp 1, 0}
        J_{\bj}^{(\bp)}(n \mp 1)
        +
        \delta_{\bj, k} \delta_{p, q} \delta_{n \mp 1, 0}
        M_{\bj}^{(p)}(n \mp 1)
        +
        \delta_{j, k} \delta_{p, q} \delta_{n \mp 2, 0}
        N_j^{(p)}(n \mp 2).
\end{aligned}
\label{eq:app_8}
\end{equation}
Shifting $n\to n\pm2$ in Eq.~\eqref{eq:app_8} and rearranging, we obtain
\begin{equation}
\begin{aligned}
        \beta_j^{(\bp)}(n)
        ={}
    &
        \frac{1}{E_j^{(\bp)}(n)}
        \Bigg[
           \beta_j^{(\bp)}(n \pm 2)
           -
           A_j^{(p)}(n \pm 2)
           \beta_j^{(p)}(n \pm 2)
           -
           F_j^{(p)}(n)
           \beta_j^{(p)}(n)
           -
           \delta_{j, k} \delta_{\bp, q} \delta_{n \pm 2, 0}
           I_j^{(\bp)}(n \pm 2)
    \\ &
           -
           \delta_{\bj, k} \delta_{\bp, q} \delta_{n \pm 1, 0}
           J_{\bj}^{(\bp)}(n \pm 1)
           -
           \delta_{\bj, k} \delta_{p, q} \delta_{n \pm 1, 0}
           M_{\bj}^{(p)}(n \pm 1)
           -
           \delta_{j, k} \delta_{p, q} \delta_{n, 0}
           N_j^{(p)}(n)
        \Bigg].
\end{aligned}
\label{eq:app_9}
\end{equation}
Substituting Eq.~\eqref{eq:app_9} into Eq.~\eqref{eq:app_7} and rearranging gives
\begin{align}
    &
        \begin{aligned}
                \beta_j^{(\bp)}(n \pm 2)
                ={}
            &
                \left(
                    F_j^{(\bp)}(n \pm 2)
                    +
                    \frac{A_j^{(\bp)}(n)}{E_j^{(\bp)}(n)}
                \right)^{-1}
                \cdot
                \Bigg[
                    \left(
                        1
                        +
                        \frac{A_j^{(\bp)}(n)}{E_j^{(\bp)}(n)}
                        F_j^{(p)}(n)
                    \right)
                    \beta_j^{(p)}(n)
            \\ &
                    -
                    \left(
                        E_j^{(p)}(n \pm 2)
                        -
                        \frac{A_j^{(\bp)}(n)}{E_j^{(\bp)}(n)}
                        A_j^{(p)}(n \pm 2)
                    \right)
                    \beta_j^{(p)}(n \pm 2)
            \\ &
                    -
                    \delta_{\bj, k} \delta_{p, q} \delta_{n \pm 1, 0}
                    \left(
                        J_{\bj}^{(p)}(n \pm 1)
                        -
                        \frac{A_j^{(\bp)}(n)}{E_j^{(\bp)}(n)}
                        M_{\bj}^{(p)}(n \pm 1)
                    \right)
            \\ &
                    -
                    \delta_{\bj, k} \delta_{\bp, q} \delta_{n \pm 1, 0}
                    \left(
                        M_{\bj}^{(\bp)}(n \pm 1)
                        -
                        \frac{A_j^{(\bp)}(n)}{E_j^{(\bp)}(n)}
                        J_{\bj}^{(\bp)}(n \pm 1)
                    \right)
            \\ &
                    -
                    \delta_{j, k} \delta_{\bp, q} \delta_{n \pm 2, 0}
                    \left(
                        N_{j}^{(\bp)}(n \pm 2)
                        -
                        \frac{A_j^{(\bp)}(n)}{E_j^{(\bp)}(n)}
                        I_j^{(\bp)}(n \pm 2)
                    \right)                    
                \Bigg]
        \end{aligned}
    \\ &
        \begin{aligned}
            =
            &
                G_j^{(p)}(n) \beta_j^{(p)}(n)
                +
                H_j^{(p)}(n \pm 2) \beta_j^{(p)}(n \pm 2)
                +
                \delta_{\bj, k} \delta_{p, q} \delta_{n \pm 1, 0}
                O_{\bj}^{(p)}(n \pm 1)
            \\ &
                +
                \delta_{\bj, k} \delta_{\bp, q} \delta_{n \pm 1, 0}
                P_{\bj}^{(\bp)}(n \pm 1)
                +
                \delta_{j, k} \delta_{\bp, q} \delta_{n \pm 2, 0}
                Q_j^{(\bp)}(n \pm 2).
        \end{aligned}
    \label{eq:app_10}
\end{align}
Applying the shift $n\to n\mp2$ to Eq.~\eqref{eq:app_10} yields
\begin{equation}
\begin{aligned}
        \beta_j^{(\bp)}(n)
        ={}
    &
        G_j^{(p)}(n \mp 2) \beta_j^{(p)}(n \mp 2)
        +
        H_j^{(p)}(n) \beta_j^{(p)}(n)
        +
        \delta_{\bj, k} \delta_{p, q} \delta_{n \mp 1, 0}
        O_{\bj}^{(p)}(n \mp 1)
    \\ &
        +
        \delta_{\bj, k} \delta_{\bp, q} \delta_{n \mp 1, 0}
        P_{\bj}^{(\bp)}(n \mp 1)
        +
        \delta_{j, k} \delta_{\bp, q} \delta_{n, 0}
        Q_j^{(\bp)}(n).
\end{aligned}
\label{eq:app_11}
\end{equation}
Substituting Eqs.~\eqref{eq:app_10} and \eqref{eq:app_11} into Eq.~\eqref{eq:app_7} gives
\begin{equation}
\begin{aligned}
        \beta_j^{(p)}(n)
        ={}
    &
        \bigg(
            E_j^{(p)}(n \pm 2)
            +
            F_j^{(\bp)}(n \pm 2)
            H_j^{(p)}(n \pm 2)
        \bigg)
        \beta_j^{(p)}(n \pm 2)
    \\ &
        +
        \bigg(
            F_j^{(\bp)}(n \pm 2)
            G_j^{(p)}(n)
            +
            A_j^{(\bp)}(n)
            H_j^{(p)}(n)
        \bigg)
        \beta_j^{(p)}(n)
    \\ &
        +
        A_j^{(\bp)}(n)
        G_j^{(p)}(n \mp 2)
        \beta_j^{(p)}(n \mp 2)
    \\ &
        +
        \delta_{\bj, k} \delta_{p, q} \delta_{n \mp 1, 0}
        A_j^{(\bp)}(n)
        O_{\bj}^{(p)}(n \mp 1)
        +
        \delta_{\bj, k} \delta_{\bp, q} \delta_{n \mp 1, 0}
        A_j^{(\bp)}(n)
        P_{\bj}^{(\bp)}(n \mp 1)
    \\ &
        +
        \delta_{j, k} \delta_{p, q} \delta_{n, 0}
        I_j^{(p)}(n)
        +
        \delta_{j, k} \delta_{p, q} \delta_{n, 0}
        A_j^{(\bp)}(n)
        Q_j^{(\bp)}(n)
    \\ &
        +
        \delta_{\bj, k} \delta_{p, q} \delta_{n \pm 1, 0}
        \left(
            F_j^{(\bp)}(n \pm 2)
            O_{\bj}^{(p)}(n \pm 1)
            +
            J_{\bj}^{(p)}(n \pm 1)
        \right)
    \\ &
        +
        \delta_{\bj, k} \delta_{\bp, q} \delta_{n \pm 1, 0}
        \left(
            F_j^{(\bp)}(n \pm 2)
            P_{\bj}^{(\bp)}(n \pm 1)
            +
            M_{\bj}^{(\bp)}(n \pm 1)
        \right)
    \\ &
        +
        \delta_{j, k} \delta_{\bp, q} \delta_{n \pm 2, 0}
        \left(
            F_j^{(\bp)}(n \pm 2)
            Q_j^{(\bp)}(n \pm 2)
            +
            N_j^{(\bp)}(n \pm 2)
        \right).
\end{aligned}
\label{eq:app_12}
\end{equation}
Restricting Eq.~\eqref{eq:app_12} to the injection channel specified by $j=k$ and $p=q$, and rearranging, we obtain
\begin{align}
    &
        \begin{aligned}
            &
                -
                \left(
                    E_j^{(p)}(n \pm 2)
                    +
                    F_j^{(\bp)}(n \pm 2)
                    H_j^{(p)}(n \pm 2)
                \right)
                \beta_j^{(p)}(n \pm 2)
            \\ &
                +
                \left(
                    1
                    -
                    F_j^{(\bp)}(n \pm 2)
                    G_j^{(p)}(n)
                    -
                    A_j^{(\bp)}(n)
                    H_j^{(p)}(n)
                \right)
                \beta_j^{(p)}(n)
            \\ &
                -
                A_j^{(\bp)}(n)
                G_j^{(p)}(n \mp 2)
                \beta_j^{(p)}(n \mp 2)
            \\ &
                =
                \delta_{n, 0}
                I_j^{(p)}(n)
        \end{aligned}
    \\ &
        \overset{\mathrm{def}}{\Longleftrightarrow}
        R_j^{(p)}(n \pm 2)
        \beta_j^{(p)}(n \pm 2)
        +
        S_j^{(p)}(n)
        \beta_j^{(p)}(n)
        +
        T_j^{(p)}(n \mp 2)
        \beta_j^{(p)}(n \mp 2)
        =
        \delta_{n, 0}
        I_j^{(p)}(n),
\end{align}
which reproduces the recurrence relation in Eq.~\eqref{eq:recurrence-formula}.

Similarly, restricting Eq.~\eqref{eq:app_11} to the injection channel specified by $j=k$ and $p=q$ gives
\begin{equation}
        \beta_j^{(\bp)}(n)
        =
        G_j^{(p)}(n \mp 2) \beta_j^{(p)}(n \mp 2)
        +
        H_j^{(p)}(n) \beta_j^{(p)}(n),
\end{equation}
which reproduces Eq.~\eqref{eq:beta-bp}.

\subsection{Explicit expressions for the auxiliary coefficients}

In Appendix~\ref{subsec:elimination_of_the_scattering_amplitudes}, we introduced several auxiliary quantities, $A_j^{(p)}(n)$, $B_j^{(p)}(n)$, $\ldots$, during  the elimination of the scattering amplitudes. Here, we derive their explicit expressions. For brevity, we use the shorthand $\delta_{j,\rL}\eqqcolon\delta$ throughout this subsection.

\begin{enumerate}[label = (\roman*)]
    \item \emph{$A$: }
        \begin{equation}
            A_j^{(p)}(n)
            =
            r_j^{(\bp)}(n \mp \delta) r_j^{(p \bp)}(n)
        \end{equation}
    \item \emph{$B$: }
        \begin{equation}
            B_j^{(p)}(n)
            =
            t_j^{(\bp)}(n \mp \delta) r_j^{(p \bp)}(n)
        \end{equation}
    \item \emph{$C$: }
        \begin{equation}
            C_j^{(p)}(n)
            =
            B_j^{(p)}(n)
            -
            \frac{A_{\bj}^{(p)}(n \mp 1)}{B_{\bj}^{(p)}(n \mp 1)}
            A_j^{(p)}(n)
            =
            \frac{r_j^{(p \bp)}(n)}{t_{\bj}^{(\bp)}(n \mp \delta)}
        \end{equation}
    \item \emph{$D$: }
        \begin{equation}
            D_j^{(p)}(n)
            =
            \frac{A_{\bj}^{(\bp)}(n \pm 1)}{B_{\bj}^{(\bp)}(n \pm 1)}
            =
            \frac{r^{(p)}_{\bj}(n \pm \delta)}{t_{\bj}^{(p)}(n \pm \delta)}
        \end{equation}
    \item \emph{$E$: }
        \begin{equation}
            E_j^{(p)}(n)
            =
            B_{\bj}^{(\bp)}(n \mp 1)
            C_j^{(p)}(n)
            =
            t_{\bj}^{(p)}\big( n \mp (2 - \delta) \big)
            r_{\bj}^{(\bp p)}(n \mp 1)
            \frac{
                r_j^{(p \bp)}(n)
            }{
                t_{\bj}^{(\bp)}(n \mp \delta)
            }
        \end{equation}
    \item \emph{$F$: }
        \begin{equation}
            F_j^{(p)}(n)
            =
            B_{\bj}^{(p)}(n \pm 1)
            D_j^{(p)}(n)
            =
            t_{\bj}^{(\bp)}\big( n \pm (2 - \delta) \big)
            r_{\bj}^{(p \bp)}(n \mp 1)
            \frac{
                r_{\bj}^{(p)}(n)
            }{
                t_{\bj}^{(p)}(n \pm \delta)
            }
        \end{equation}
    \item \emph{$G$: }
        \begin{align}
        \begin{aligned}
                G_j^{(p)}(n)
            &
                =
                \left(
                    F_j^{(\bp)}(n \pm 2)
                    +
                    \frac{
                        A_j^{(\bp)}(n)
                    }{
                        E_j^{(\bp)}(n)
                    }
                \right)^{-1}
                \cdot
                \left(
                    1
                    +
                    \frac{
                        A_j^{(\bp)}(n)
                    }{
                        E_j^{(\bp)}(n)
                    }
                    F_j^{(p)}(n)
                \right)
            \\ &
                =
                \frac{
                    t_{\bj}^{(\bp)}\big( n \pm (2 - \delta) \big)
                    r_{\bj}^{(p \bp)}(n \pm 1)
                    \cT(n \pm \delta)
                }{
                    t_{\bj}^{(p)}(n \pm \delta)
                    \Big(
                        r_{\bj}^{(p \bp)}(n \pm 1)
                        r_{\bj}^{(\bp p)}(n \pm 1)
                        r_{\bj}^{(\bp)}\big( n \pm (2 - \delta) \big)
                        +
                        r_j^{(p)}(n \pm \delta)
                    \Big)
                }
        \end{aligned}
        \end{align}
    \item \emph{$H$: }
        \begin{align}
        \begin{aligned}
                H_j^{(p)}(n)
            &
                =
                -\left(
                    F_j^{(\bp)}(n)
                    +
                    \frac{
                        A_j^{(\bp)}(n \mp 2)
                    }{
                        E_j^{(\bp)}(n \mp 2)
                    }
                \right)^{-1}
                \cdot
                \left(
                    E_j^{(p)}(n)
                    -
                    \frac{
                        A_j^{(\bp)}(n \mp 2)
                    }{
                        E_j^{(\bp)}(n \mp 2)
                    }
                    A_j^{(p)}(n)
                \right)
            \\ &
                =
                \frac{
                    -r_j^{(p \bp)}(n)
                    \Big(
                        r_{\bj}^{(p \bp)}(n \mp 1)
                        r_{\bj}^{(\bp p)}(n \mp 1)
                        -
                        r_j^{(\bp)}(n \mp \delta)
                        r_j^{(p)}\big( n \mp (2 - \delta) \big)
                    \Big)
                }{
                    r_{\bj}^{(p \bp)}(n \mp 1)
                    r_{\bj}^{(\bp p)}(n \mp 1)
                    r_{\bj}^{(\bp)}(n \mp \delta)
                    +
                    r_j^{(p)}\big( n \mp (2 - \delta) \big)
                }
        \end{aligned}
        \end{align}
    \item \emph{$R$: }
        \begin{align}
        \begin{aligned}
                R_j^{(p)}(n \pm 2)
            &
                =
                -E_j^{(p)}(n \pm 2)
                -
                F_j^{(\bp)}(n \pm 2)
                H_j^{(p)}(n \pm 2)
            \\ &
                =
                \frac{
                    -r_j^{(p)}(n \pm \delta)
                    t_{\bj}^{(p)}(n \pm \delta)
                    r_{\bj}^{(\bp p)}(n \pm 1)
                    r_j^{(p \bp)}(n \pm 2)
                    \cT\big( n \pm (2 - \delta) \big)
                }{
                    t_{\bj}^{(\bp)}\big( n \pm (2 - \delta) \big)
                    \Big(
                        r_{\bj}^{(p \bp)}(n \pm 1)
                        r_{\bj}^{(\bp p)}(n \pm 1)
                        r_{\bj}^{(\bp)}\big( n \pm (2 - \delta) \big)
                        +
                        r_j^{(p)}(n \pm \delta)
                    \Big)
                }
        \end{aligned}
        \end{align}
    \item \emph{$S$: }
        \begin{align}
        \begin{alignedat}{3}
                S_j^{(p)}(n)
            &
                ={}
            & &
                1
                -
                F_j^{(\bp)}(n \pm 2)
                G_j^{(p)}(n)
                -
                A_j^{(\bp)}(n)
                H_j^{(p)}(n)
            \\ &
                ={}
            & &
                1
                -
                \frac{
                    r_{\bj}^{(p \bp)}(n \pm 1)
                    r_{\bj}^{(\bp p)}(n \pm 1)
                    r_{\bj}^{(\bp)}\big( n \pm (2 - \delta) \big)
                    \cT(n \pm \delta)
                }{
                    r_{\bj}^{(p \bp)}(n \pm 1)
                    r_{\bj}^{(\bp p)}(n \pm 1)
                    r_{\bj}^{(\bp)}\big( n \pm (2 - \delta) \big)
                    +
                    r_j^{(p)}(n \pm \delta)
                }
            \\ & & &
                +
                r_j^{(p)}(n \pm \delta)
                r_j^{(\bp p)}(n)
                \cdot
                \frac{
                    r_j^{(p \bp)}(n)
                    \Big(
                        r_{\bj}^{(p \bp)}(n \mp 1)
                        r_{\bj}^{(\bp p)}(n \mp 1)
                        -
                        r_j^{(\bp)}(n \mp \delta)
                        r_j^{(p)}\big( n \mp (2 - \delta) \big)
                    \Big)
                }{
                    r_{\bj}^{(p \bp)}(n \mp 1)
                    r_{\bj}^{(\bp p)}(n \mp 1)
                    r_{\bj}^{(\bp)}(n \mp \delta)
                    +
                    r_j^{(p)}\big( n \mp (2 - \delta) \big)
                }
        \end{alignedat}
        \end{align}
    \item \emph{$T$: }
        \begin{align}
        \begin{aligned}
                T_j^{(p)}(n \mp 2)
            &
                =
                -A_j^{(\bp)}(n)
                G_j^{(p)}(n \mp 2)
            \\ &
                =
                \frac{
                    -
                    r_j^{(p)}(n \pm \delta)
                    r_j^{(\bp p)}(n)
                    t_{\bj}^{(\bp)}(n \mp \delta)
                    r_{\bj}^{(p \bp)}(n \mp 1)
                    \cT\big( n \mp (2 - \delta) \big)
                }{
                    t_{\bj}^{(p)}\big( n \mp (2 - \delta) \big)
                    \Big(
                        r_{\bj}^{(p \bp)}(n \mp 1)
                        r_{\bj}^{(\bp p)}(n \mp 1)
                        r_{\bj}^{(\bp)}(n \mp \delta)
                        +
                        r_j^{(p)}\big( n \mp (2 - \delta) \big)
                    \Big)
                }
        \end{aligned}
        \end{align}
\end{enumerate}

Apart from differences in notation, the coefficients $R_j^{(p)}(n \pm 2)$, $S_j^{(p)}(n)$, and $T_j^{(p)}(n \mp 2)$ agree with the corresponding expressions in Ref.~\cite{Hofmann}.

\section{Numerical solution of the recurrence relation}
\label{app:numerical_solution_of_the_recurrence_relation}

In general, the recurrence relation~\eqref{eq:recurrence-formula} does not admit a closed-form analytical solution for $\beta_j^{(p)}$. To solve it numerically,
 we write Eq.~\eqref{eq:recurrence-formula} in matrix form using an $(N + 1)\times(N + 1)$ tridiagonal matrix $\mathbf{M}_{j}^{(p)}(N)$:
\begin{equation}
    \lim_{N \to \infty}
        \mathbf{I}_{j}^{(p)}(N)
    =
    \lim_{N \to \infty}
        \mathbf{M}_{j}^{(p)}(N) \bm{\beta}_{j}^{(p)}(N).
\end{equation}

According to the notation for the sideband indices introduced in Eqs.~\eqref{eq:sideband_index_notation} and \eqref{eq:sideband_index_rules}, the two cases $(j, p)=(\rL, \re), (\rR, \rh)$ and $(j, p)=(\rL, \rh), (\rR, \re)$ must be treated separately. The corresponding vectors and matrices are defined explicitly as follows:
\begin{enumerate}[label = (\roman*)]
    \item \emph{For $(j, p) = (\rL, \re), (\rR, \rh)$:}
        \begin{subequations}
        \begin{gather}
                \mathbf{I}_{j}^{(p)}(N)
                =
                \begin{bmatrix}
                       0
                    &
                        \cdots
                    &
                        I_{j}^{(p)}(0)
                    &
                        \cdots
                    &
                        0
    			\end{bmatrix}^\mathrm{T}
                ,
            \\
                \bm{\beta}_{j}^{(p)}(N)
                =
                \begin{bmatrix}
    				    \beta_j^{(p)}(N)
                    &
                        \beta_j^{(p)}(N - 2)
                    &
                        \cdots
                    &
                        \beta_j^{(p)}(2)
                    &
                        \beta_j^{(p)}(0)
                    &
                        \beta_j^{(p)}(-2)
                    &
                        \cdots
                    &
                        \beta_j^{(p)}(-N)
    			\end{bmatrix}^\mathrm{T}
                ,
            \\
                \begin{aligned}
                    &
                        \mathbf{M}_{j}^{(p)}(N)
                    \\ &
                        \coloneqq
                        \begin{bmatrix}
            					S_{j}^{(p)}(N)
                            &
                                T_{j}^{(p)}(N - 2)
                            &
                                0
                            &
                                \cdots
                            &&&&&&
            				\\
            					R_{j}^{(p)}(N)
                            &
                                S_{j}^{(p)}(N - 2)
                            &
                                T_{j}^{(p)}(N - 4)
                            &
                                0
                            &
                                \cdots
                            &&&&&
            				\\
                            &
                                \vdots
                            &&&&
                                \vdots
                            &&&
            				\\
                            &&
                                \cdots
                            &
                                0
                            &
                                R_{j}^{(p)}(2)
                            &
                                S_{j}^{(p)}(0)
                            &
                                T_{j}^{(p)}(-2)
                            &
                                0
                            &
                                \cdots 
            				\\
                            &&&&&
                                \vdots
                            &&&&
                                \vdots
            				\\
                            &&&&&&
                                \cdots
                            &
                                0
                            &
                                R_{j}^{(p)}(-N + 2)
                            &
                                S_j^{(p)}(-N)
            			\end{bmatrix}
                        .
                \end{aligned}
        \end{gather}
        \end{subequations}
    \item \emph{For $(j, p) = (\rL, \rh), (\rR, \re)$:}
        \begin{subequations}
        \begin{gather}
                \mathbf{I}_{j}^{(p)}(N)
                =
                \begin{bmatrix}
                       0
                    &
                        \cdots
                    &
                        I_{j}^{(p)}(0)
                    &
                        \cdots
                    &
                        0
    			\end{bmatrix}^\mathrm{T}
                ,
            \\
                \bm{\beta}_{j}^{(p)}(N)
            =
                \begin{bmatrix}
    				    \beta_j^{(p)}(-N)
                    &
                        \beta_j^{(p)}(-N + 2)
                    &
                        \cdots
                    &
                        \beta_j^{(p)}(-2)
                    &
                        \beta_j^{(p)}(0)
                    &
                        \beta_j^{(p)}(2)
                    &
                        \cdots
                    &
                        \beta_j^{(p)}(N)
    			\end{bmatrix}^\mathrm{T}
                ,
            \\
                \begin{aligned}
                    &
                        \mathbf{M}_{j}^{(p)}(N)
                    \\ &
                        =
                        \begin{bmatrix}
            					S_{j}^{(p)}(-N)
                            &
                                T_{j}^{(p)}(-N + 2)
                            &
                                0
                            &
                                \cdots
                            &&&&&&
            				\\
            					R_{j}^{(p)}(-N)
                            &
                                S_{j}^{(p)}(-N + 2)
                            &
                                T_{j}^{(p)}(-N + 4)
                            &
                                0
                            &
                                \cdots
                            &&&&&
            				\\
                            &
                                \vdots
                            &&&&
                                \vdots
                            &&&
            				\\
                            &&
                                \cdots
                            &
                                0
                            &
                                R_{j}^{(p)}(-2)
                            &
                                S_{j}^{(p)}(0)
                            &
                                T_{j}^{(p)}(2)
                            &
                                0
                            &
                                \cdots 
            				\\
                            &&&&&
                                \vdots
                            &&&&
                                \vdots
            				\\
                            &&&&&&
                                \cdots
                            &
                                0
                            &
                                R_{j}^{(p)}(N - 2)
                            &
                                S_j^{(p)}(N)
            			\end{bmatrix}
                        .
                \end{aligned}
        \end{gather}
        \end{subequations}
\end{enumerate}

Since the limit $N\to\infty$ cannot be implemented directly in a numerical calculation, we instead choose a sufficiently large finite value of $N$ and determine $\bm{\beta}_{j}^{(p)}(N)$ numerically from the inverse matrix ${\mathbf{M}_{j}^{(p)}(N)}^{-1}$. This truncation provides an accurate approximation because the Andreev-reflection probability approaches zero in the large-$N$ limit.

\section{Derivation of electric and heat currents}
\label{app:derivation_currents}

\subsection{Wave-function representation}
With reference to the scattering-state representation in Ref.~\cite{Hofmann}, the wave function $\Psi_k^{(q)}(E,x,t)$ appearing in the field-operator expansion~\eqref{eq:field-operator_expansion} can be written as 
\begin{align}
\begin{aligned}
    &
        \Psi_k^{(q)}(E, x, t)
    \\ &
        =
        \sqrt{\frac{m}{\hbar}}
        \sum_{n = -\infty}^{\infty}
        \sum_{\gamma = \alpha, \beta}
        \begin{bmatrix}
                \exp{\Big(
                    -\frac{i}{\hbar} \big( E + (n + \delta_{k, \rL}) V \big) t
                \Big)}
                \cdot
                \frac{
                    \gamma_k^{(\re q)}(E, n - \delta_{k, \rL})
                }{
                    \sqrt{q_\re(E, n - \delta_{k, \rL})}
                }
                \cdot
                \exp{\big(
                    \zeta_{\gamma \alpha}
                    \zeta_{k \rL}
                    i q_\re(E, n - \delta_{k, \rL}) x
                \big)}
            \\
                \exp{\Big(
                    -\frac{i}{\hbar} \big( E + (n - \delta_{k, \rL}) V \big) t
                \Big)}
                \cdot
                \frac{
                    \gamma_k^{(\rh q)}(E, n + \delta_{k, \rL})
                }{
                    \sqrt{q_\rh(E, n + \delta_{k, \rL})}
                }
                \cdot
                \exp{\big(
                    -\zeta_{\gamma \alpha}
                    \zeta_{k \rL}
                    i q_\rh(E, n + \delta_{k, \rL}) x
                \big)}
        \end{bmatrix},
\end{aligned}
\label{eq:wave-function}
\end{align}
where $q_p(E, n)$ is defined by
\begin{align}
    q_p(E, n)
    \coloneqq
    \sqrt{
        \frac{2m}{\hbar^2}
        \big(
            \mu_\rR + \zeta_{p \re} (E + nV)
        \big)
    }.
\end{align}

\subsection{Derivation of the electric current}
Substituting the field-operator expansions~\eqref{eq:field-operator_expansion} and \eqref{eq:wave-function} into the definition of the electric current~\eqref{eq:def_electric-current}, we obtain

\begin{align}
\begin{aligned}
    &
        J_\mathrm{ele}(x, t)
    \\ &
        \begin{aligned}
                ={}
            &
                \frac{e}{2i}
                \cdot
                \frac{1}{h v_\rF}
                \sum_{k, q}
                \int_{-\infty}^{\infty} \rd E \
                D(E - \zeta_{q \re} \delta_{k, \rL} V)
                \sum_{n, n'} \sum_{\gamma, \gamma'}
            \\ &
                \Bigg[
                    \big(
                        \zeta_{\gamma' \alpha}
                        \zeta_{k \rL}
                        i q_\re(E, n' - \delta_{k, \rL})
                        +
                        \zeta_{\gamma \alpha}
                        \zeta_{k \rL}
                        i q_\re(E, n - \delta_{k, \rL})
                    \big)
            \\ &
                    \cdot
                    \exp{\bigg(
                        -\frac{i}{\hbar}
                        \Big(
                            \big( E + (n' + \delta_{k, \rL})V \big)
                            -
                            \big( E + (n + \delta_{k, \rL})V \big)
                        \Big)
                        t
                    \bigg)}
                    \cdot
                    \frac{
                        {\gamma^{(\re q)}_k(E, n - \delta_{k, \rL})}^*
                        {\gamma'}^{(\re q)}_k(E, n' - \delta_{k, \rL})
                    }{
                        \sqrt{q_\re(E, n - \delta_{k, \rL}) q_\re(E, n' - \delta_{k, \rL})}
                    }
            \\ &
                    \cdot
                    \exp{\Big(
                        \zeta_{k \rL} i
                        \big(
                            \zeta_{\gamma' \alpha} q_\re(E, n' - \delta_{k, \rL})
                            -
                            \zeta_{\gamma \alpha} q_\re(E, n - \delta_{k, \rL})
                        \big)
                        x
                    \Big)}
                    \cdot
                    \left\langle
                        \hat{a}_{k \sigma_q}^{(\bq)}(\zeta_{q \re} E - \delta_{k, \rL} V)
                        \hat{a}_{k \sigma_q}^{(q)}(\zeta_{q \re} E - \delta_{k, \rL} V)
                    \right\rangle
            \\ &
                    +
                    \big(
                        \zeta_{\gamma' \alpha}
                        \zeta_{k \rL}
                        i q_\rh(E, n' + \delta_{k, \rL})
                        +
                        \zeta_{\gamma \alpha}
                        \zeta_{k \rL}
                        i q_\rh(E, n + \delta_{k, \rL})
                    \big)
            \\ &
                    \cdot
                    \exp{\bigg(
                        -\frac{i}{\hbar}
                        \Big(
                            \big( E + (n' - \delta_{k, \rL})V \big)
                            -
                            \big( E + (n - \delta_{k, \rL})V \big)
                        \Big)
                        t
                    \bigg)}
                    \cdot
                    \frac{
                        \gamma^{(\rh q)}_k(E, n + \delta_{k, \rL})
                        {{\gamma'}^{(\rh q)}_k(E, n' + \delta_{k, \rL})}^*
                    }{
                        \sqrt{q_\rh(E, n + \delta_{k, \rL}) q_\rh(E, n' + \delta_{k, \rL})}
                    }
            \\ &
                    \cdot
                    \exp{\Big(
                        \zeta_{k \rL} i
                        \big(
                            \zeta_{\gamma' \alpha} q_\rh(E, n' + \delta_{k, \rL})
                            -
                            \zeta_{\gamma \alpha} q_\rh(E, n + \delta_{k, \rL})
                        \big)
                        x
                    \Big)}
                    \cdot
                    \left\langle
                        \hat{a}_{k \sigma_q}^{(q)}(\zeta_{q \re} E - \delta_{k \rL} V)
                        \hat{a}_{k \sigma_q}^{(\bq)}(\zeta_{q \re} E - \delta_{k \rL} V)
                    \right\rangle
                \Bigg].
        \end{aligned}        
\end{aligned}
\end{align}
Since we are interested in the time-independent dc component, we retain only the terms satisfying
\begin{equation}
    \exp{\bigg(
        -\frac{i}{\hbar}
        \Big(
            \big( E + (n' \pm \delta_{k, \rL})V \big)
            -
            \big( E + (n \pm \delta_{k, \rL})V \big)
        \Big)
        t
    \bigg)}
    =
    1
    \quad
    \therefore n = n',
\end{equation}
which yields Eq.~\eqref{eq:electric-current}.

\subsection{Derivation of the heat current}
Inspection of the heat-current expression~\eqref{eq:formalism_heat-current} shows that its second term is obtained directly from the definition of the electric current~\eqref{eq:def_electric-current} by replacing $e \to -\mu_k$. The first term likewise has the same structure as the electric-current expression, apart from a factor of $q_p(E, n - \zeta_{p \re} \delta_{k \rL})^2$ and the corresponding prefactors. Therefore, the heat-current expression is obtained from Eq.~\eqref{eq:electric-current} by making the replacement
\begin{equation}
    \frac{e}{h v_\rF}
    \to
    \frac{-\mu_k}{h v_\rF}
    +
    \frac{\hbar q_p(E, n - \zeta_{p \re} \delta_{k \rL})^2}{4 \pi m v_\rF}
    =
    \frac{\zeta_{p \re}}{h v_\rF}
    (E + nV),
\end{equation}
which yields Eq.~\eqref{eq:heat-current}.

\twocolumngrid

\clearpage
\bibliography{references}

\end{document}